\documentclass[acmsmall,natbib=false]{acmart}
\usepackage{mathtools}
\usepackage[ruled,linesnumbered]{algorithm2e}

\usepackage{amssymb,bm}
\usepackage{amsmath}
\DeclareMathOperator*{\argmax}{argmax} 
\DeclareMathOperator*{\argmin}{argmin} 
\usepackage{pifont}% http://ctan.org/pkg/pifont
\usepackage{color,xcolor}
\usepackage{multirow}
\usepackage{listings}
\usepackage[shortlabels]{enumitem}
\usepackage{tcolorbox} 
\usepackage{xfrac}

\usepackage{array}
\newcolumntype{?}{!{\vrule width 1pt}}

\usepackage{tikz}

\definecolor{dkgreen}{rgb}{0,0.6,0}
\definecolor{ltgray}{rgb}{0.5,0.5,0.5}

\definecolor{pink}{rgb}{1.0, 0.74, 0.85}

\newcommand{\cmark}{\ding{51}}%
\newcommand{\xmark}{\ding{55}}%
\newcommand{\parh}[1]{\vspace{1.5pt}\noindent\textbf{#1}}

\makeatletter
\newcommand{\thickhline}{%
    \noalign {\ifnum 0=`}\fi \hrule height 1pt
    \futurelet \reserved@a \@xhline
}
\makeatother

\newtheorem{definition}{Definition}[section]
\newtheorem{theorem}{Theorem}[section]

\newtheorem{lemma}{Lemma}[subsection]
\newtheorem{proposition}{Proposition}[section]

\newtheoremstyle{ugh}
{3pt}% hSpace above
{3pt}% hSpace below
{}% hBody font
{}% hIndent amount
{\bfseries}% hTheorem head font
{:}% hPunctuation after theorem head
{.5em}% hSpace after theorem head
{}% hTheorem head spec (can be left empty, meaning ‘normal’)

\theoremstyle{ugh}
\newtheorem{eexample}{Example}[section]

\newenvironment{changemargin}{%
\begin{list}{}{%
\setlength{\topsep}{0pt}%
\setlength{\leftmargin}{0pt}%    
\setlength{\rightmargin}{0pt}%  
\setlength{\listparindent}{\parindent}%
\setlength{\itemindent}{\parindent}%
\setlength{\parsep}{\parskip}%
}%
\item[]}{\end{list}}

\newenvironment{example}{%
\begin{changemargin}\vskip-\baselineskip
\begin{eexample}}{%
\end{eexample}\end{changemargin}}

\newtcolorbox{rqbox}[3][]
{
  colframe = rq,
  colback  = rqBack,
  coltitle = rq!10,  
  title    = {#3},
  fontupper=\itshape,
  #1,
}

\definecolor{rq}{HTML}{1B365C}
\definecolor{rqBack}{HTML}{9ECBF7}

\newcommand{\tool}{\textsc{XInsight}}
\newcommand{\xlearn}{\textsc{XLearner}}
\newcommand{\xplainer}{\textsc{XPlainer}}
\newcommand{\xtrans}{\textsc{XTranslator}}

\newcommand{\diff}{\textsc{Why Query}}
\newcommand{\tuple}[1]{\ensuremath{\left \langle #1 \right \rangle }}

\newcommand{\revision}[1]{#1}

\newcommand{\fdarrow}{\xrightarrow{\texttt{FD}}}
\newcommand{\fdgraph}{\mathcal{G}_{\texttt{FD}}}
\newcommand{\xfd}{\bm{X}_{\texttt{FD}}}
\newcommand{\sm}{Supplementary Material}

\newcommand{\F}{Fig.}
\newcommand{\E}{Eqn.}
\newcommand{\Ex}{Ex.}
\newcommand{\T}{Table}
\renewcommand{\S}{Sec.}
\newcommand{\A}{Alg.}
\newcommand{\D}{Def.}
\newcommand{\Lem}{Lemma.}
\newcommand{\Thm}{Thm.}

\newcommand{\Prop}{Proposition}
\newcommand{\ignore}[1]{}
 
\def\sep{{\Perp_{\mathcal{G}}}}

\def\circarrow{{\circ\hspace{0.3mm}\!\!\! \rightarrow}}
\def\stararrow{{*\hspace{0.3mm}\!\!\! \rightarrow}}
\def\starlinestar{{*\hspace{0.3mm}\!\!\! - \hspace{0.3mm}\!\!\!*}}
\def\circlinecirc{{\circ \hspace{0.4mm}\!\!\! - \hspace{0.4mm}\!\!\!\circ}}
\def\circline{{\circ \! -}}
\def\circstar{{\circ \! - \hspace{0.3mm}\!\!\!*}}
\def\starcirc{{*\hspace{0.3mm}\!\!\! - \hspace{0.3mm}\!\!\! \circ}}
\def\arrowcirc{{\leftarrow\hspace{0.3mm}\!\!\! \circ}}
\def\arrowstar{{\leftarrow\hspace{0.3mm}\!\!\!*}}
\def\linestar{{-\hspace{0.3mm}\!\!\!*}}
\def\linecirc{{- \hspace{0.3mm}\!\!\! \circ}}
\AtBeginDocument{%
  }

\setcopyright{acmlicensed}
\acmJournal{PACMMOD}
\acmYear{2023} \acmVolume{1} \acmNumber{2} \acmArticle{156} \acmMonth{6} \acmPrice{15.00}\acmDOI{10.1145/3589301}

\received{October 2022}
\received[revised]{January 2023}
\received[accepted]{February 2023}

\RequirePackage[
  datamodel=acmdatamodel,
  ]{biblatex}

\begin{document}

\title{\tool: eXplainable Data Analysis Through The Lens of Causality}

\author{Pingchuan Ma}
\affiliation{%
  \institution{Hong Kong University of Science and Technology}
  \city{Kowloon}
  \country{Hong Kong SAR}
}
\email{pmaab@cse.ust.hk}

\author{Rui Ding}\authornote{Corresponding author.}
\affiliation{%
  \institution{Microsoft Research}
  \city{Beijing}
  \country{China}
}
\email{juding@microsoft.com}

\author{Shuai Wang}
\affiliation{%
\institution{Hong Kong University of Science and Technology}
\city{Kowloon}
  \country{Hong Kong SAR}
}
\email{shuaiw@cse.ust.hk}

\author{Shi Han}
\affiliation{%
  \institution{Microsoft Research}
  \city{Beijing}
  \country{China}
}

\author{Dongmei Zhang}
\affiliation{%
  \institution{Microsoft Research}
  \city{Beijing}
  \country{China}
}

\renewcommand{\shortauthors}{Pingchuan Ma et al.}

\begin{abstract}
  In light of the growing popularity of Exploratory Data Analysis (EDA),
  understanding the underlying causes of the knowledge acquired by EDA is
  crucial. However, it remains under-researched. This study promotes a
  transparent and explicable perspective on data analysis, called
  \textit{eXplainable Data Analysis} (XDA). For this reason, we present \tool, a
  general framework for XDA. \tool\ provides data analysis with qualitative and
  quantitative explanations of causal and non-causal semantics. This way, it
  will significantly improve human understanding and confidence in the outcomes
  of data analysis, facilitating accurate data interpretation and decision
  making in the real world. \tool\ is a three-module, end-to-end pipeline
  designed to extract causal graphs, translate causal primitives into XDA
  semantics, and quantify the quantitative contribution of each explanation to a
  data fact. \tool\ uses a set of design concepts and optimizations to address
  the inherent difficulties associated with integrating causality into XDA.
  Experiments on synthetic and real-world datasets as well as a user study
  demonstrate the highly promising capabilities of \tool. 
  % \tool\ has been partially integrated into Microsoft Power BI.
\end{abstract}



\maketitle

\section{Introduction}
\label{sec:introduction}

Exploratory data analysis (EDA) is key to acquiring insight from data and
facilitating analysis towards decision
making~\cite{milo2020automating,ma2021metainsight}. With the advent of the
digital age, the information explosion phenomenon~\cite{buckland2017information}
makes it difficult for users to justify and rely on knowledge and conclusions
from EDA. To ease the cognitive process, data explanations are proposed to
deliberate data facts (e.g., query outcomes) and enhance user
comprehension~\cite{glavic2021trends}. In this paper, we term such a process as
\textit{eXplainable Data Analysis} (XDA), which advances data analysis by
providing users with effective explanations. \revision{By suggesting and
justifying choices to alter outcomes, XDA helps users comprehend and trust
phenomena emerging from data;} as a result, it facilitates real-world decision
making.

\begin{figure*}[!htbp]
\centering
\includegraphics[width=\textwidth]{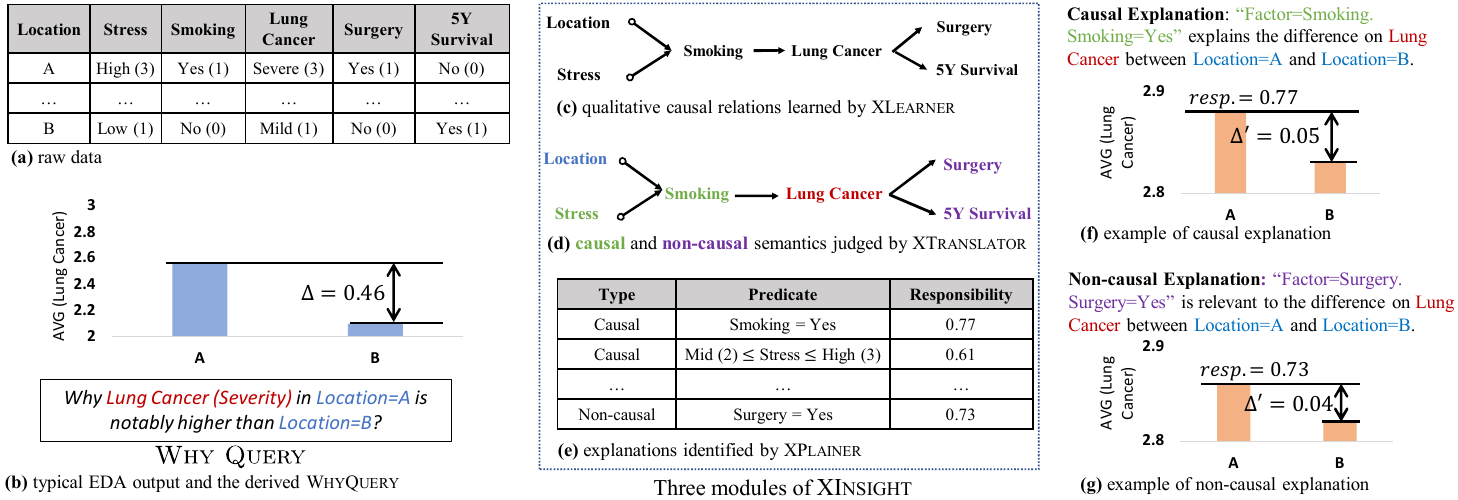}
\vspace{-8pt}
\caption{Illustrative example of \tool.}
\label{fig:motivating-example}
\vspace{-10pt}
\end{figure*}

Explanations can be categorized as either causal or
non-causal~\cite{lange2016because}.
%, whereas the former one is generally more
%\fm{expressive}.
%
Causal explanations seek causal factors to explain an outcome.
\F~\ref{fig:motivating-example} depicts a hypothetical lung cancer dataset.
\revision{Here, a patient's location (indicating regional tobacco control
policy) and amount of stress have an impact on whether they would smoke. Then,
smoking influences lung cancer's severity.} The degree of severity further
affects whether they would undergo surgery and the five-year survival rate.
Here, smoking explains why a patient has high lung cancer severity (see
\F~\ref{fig:motivating-example}(f)). 
In contrast, a non-causal explanation shows the results merely by statistical
correlations. For example, surgery ``explains'' (more precisely, is relevant to)
lung cancer severity (see \F~\ref{fig:motivating-example}(g)). Despite being
helpful, this is not a causal explanation~\cite{povich2018because}.

%Despite the prosperous progresses have been made by the database community, 
Existing data explanation tools (e.g., Tableau's Explain Data~\cite{tableau} in
industry, Scorpion~\cite{wu2013scorpion} and DIFF~\cite{abuzaid2021diff} in
academia) often provide non-causal explanations~\cite{glavic2021trends}.
Although valuable for data analysis, they may mislead users who want causal
explanations. A well-known confusion, as noted in~\cite{law2021causal}, is that
Tableau's Explain Data reports that Massachusetts' low teenage pregnancy rate
may explain this state's high ACT Math score. Such explanations are
questionable. In comparison, causal explanations play a central role in human
cognition~\cite{keil2006explanation,murphy1985role}. They enable users to make
counterfactual thinking and actionable decisions. \revision{For instance,
quitting smoking reduces lung cancer severity whereas cancelling surgery does
not.}

According to Pearl's general causality~\cite{judea2009causal}, causal knowledge
is typically represented by causal graphs. Each node in a causal graph
represents a random variable in data, and each directed edge between parent and
child nodes denotes a cause-effect relation. Causal knowledge primarily conveys
\textit{qualitative explanations}~\cite{scheines1991qualitative}, such as
smoking causes lung cancer. To further enable \textit{quantitative
explanations}, \revision{it is necessary to quantify the contribution of each
input to the output. This way, we quantify smoking's contribution to lung cancer
and compare it with other factors. Halpern's actual
causality~\cite{halpern2005causes,halpern2016actual}, and essentially its
adaptation, DB Causality [30], provide an elegant formulation of this concept.}

This paper proposes \tool\ as a unified, causality-based XDA framework that
qualitatively and quantitatively answers \diff\ raised by users. Considering the
following \diff:

\begin{example}
    The dataset in \F~\ref{fig:motivating-example}(a) depicts the patient
    information in a country. An analyst observes an interesting data fact:
    ``the average severity of lung cancer of patients in location A is much
    higher than in location B'' and then raises a \diff\
    (\F~\ref{fig:motivating-example}(b)). 
\end{example}

\noindent for which \F~\ref{fig:motivating-example}(f) and (g) illustrate two
explanations provided by \tool. Each explanation is flagged as either a
``causal'' or ``non-causal'' explanation, and is composed of both qualitative
and quantitative sub-explanations.

\begin{example}
    An explanation (\F~\ref{fig:motivating-example}(f)) deems ``Smoking'' to be
    a qualitative causal factor of lung cancer severity and highlights
    ``Smoking=Yes'' and its responsibility as a quantitative sub-explanation.
\end{example}

\tool\ includes three modules, \xlearn, \xtrans, and \xplainer\ to gradually
form explanations. \xlearn\ first automatically discovers a causal graph
$\mathcal{G}$ from data (\F~\ref{fig:motivating-example}(c)). Then, given a
\diff\ (\F~\ref{fig:motivating-example}(b)) with the target (i.e., the measure
``Lung Cancer'') and the {context} (i.e., the breakdown dimension 
% ;\revision{see the definition of context in \S~\ref{subsec:prel-data}}
``Location''), \xtrans\ enumerates each remaining variable on $\mathcal{G}$ and
decides if it is causal or non-causal to the target under the context.
\F~\ref{fig:motivating-example}(d) shows causal variables (i.e., those that can
potentially provide causal explanations) in green and non-causal variables
(i.e., those that can potentially provide non-causal explanations) in purple.
Last, \xplainer\ quantifies how well each variable answers \diff\ by searching
possible predicates on the variables that are the most responsible, as shown in
\F~\ref{fig:motivating-example}(e). 
% We concretize \tool\ for explaining differences in data, which is a typical
% scenario and also extensively studied in previous
% works~\cite{abuzaid2021diff,wu2013scorpion,glavic2021trends}.
Despite the promising capability of \tool, concretizing each module is
challenging. We brief the challenges and our solutions in the following.

\parh{\xlearn.}~Most real-world datasets are collected irrespective of causal
sufficiency~\cite{peters2017elements}. In other words, not all causally relevant
variables are available in the dataset. Furthermore, real-world datasets often
contain deterministic relations in the form of Functional Dependency (FD),
especially when they have materialized from relational databases. These FDs may
violate the faithfulness assumption~\cite{ding2020reliable}, which is crucial
for many causal discovery algorithms. To address these challenges, we establish
a theory to propose an FD-induced graph $\fdgraph$. \xlearn\ uses $\fdgraph$ to
select a subset of variables for standard causal discovery where the selected
variables do not trigger faithfulness violations induced by FDs. It adopts
FCI~\cite{zhang2008completeness} to address causal insufficiency and
synergistically combine the result of FCI with the causal relations entailed by
$\fdgraph$.

\parh{\xtrans.}~The translation from causal primitives (the structural relations
in the causal graph) into XDA semantics (e.g., whether a variable provides
causal explanations) 
%  or has no explanation power
% \revision{see details in \T~\ref{tab:pag-paths}}
is under-explored. Given a \diff\ (with a target and a context), it is unclear
how to determine if a variable $X$ can explain the target given the context,
and, moreover, if $X$ provides causal or non-causal explanations. \xtrans\
characterizes various causal primitives (e.g., m-separation, ancestor/descendant
relations) from a causal graph and provides a taxonomy to translate them into
XDA semantics.

\parh{\xplainer.}~DB causality is primarily designed for data provenance, which
usually provides tuples as explanations. Contrarily, we note that predicate-level
explanations shall be more desirable for XDA scenarios. Moreover, computing the
responsibility of explanations with DB causality is NP-complete in
general~\cite{meliou2010complexity}. \xplainer\ adapts DB causality to XDA by
using predicate-level explanations with the conciseness consideration and also
significantly reduces the computing cost with theoretical guarantees.
In summary, we make the following contributions:

\begin{itemize}[leftmargin=3mm]
   \item We propose \tool, a unified and causality-based framework for XDA.
   XInsight features adequate (by distinguishing causal from non-causal) and
   comprehensive (with qualitative and quantitative) explanations.
    \item \tool\ consists of three modules, \xlearn, \xtrans\ and \xplainer,
    each of which is meticulously designed to address technical challenges and
    deliver efficient analysis. \xlearn\ learns the causal graph from causally
    insufficient data in the presence of FD-induced faithfulness violations,
    \xtrans\ translates causal primitives into XDA semantics, and \xplainer\
    efficiently provides quantitative explanations via an adaptation of DB
    causality to meet the needs of XDA scenarios.
    \item Empirically, we conduct thorough experiments on public data,
    production data, and synthetic data via quantitative experiments and human
    evaluations. The results are very encouraging.
\end{itemize}

\parh{\revision{Open Source and Real-world Adoption.}}~We release our code
at~\cite{code}. \xplainer\ has been integrated into Microsoft Power BI to
explain increase/decrease in data~\cite{pbi}.

\section{Preliminary}
\label{sec:preliminary}

\subsection{Data Model and Query}
\label{subsec:prel-data}

\noindent\textbf{Multi-Dimensional Data.}~Let $D\coloneqq\{X_1, \cdots,X_n\}$
represents multi-dimensional data comprising $n$ attributes. In \tool, we assume
that records of $D$ are drawn independently from an identical distribution
without selection biases (i.e, i.i.d.~assumption) such that each attribute is a
(random) variable. Here, selection bias is a preferential selection of units in
data analysis~\cite{bareinboim2012controlling}. A variable is either categorical
or numerical. In accordance with previous
works~\cite{ma2021metainsight,ding2019quickinsights}, we denote a categorical
variable as \textit{dimension} and a numerical variable as \textit{measure}.
% Categorical attributes are denoted as dimension
% and numerical attributes are denoted as measure. 
\revision{Multi-dimensional data is commonly represented as a spreadsheet in our
context. For relational data, we anticipate taking a materialized provenance
table~\cite{li2021putting} as input.}

\noindent\textbf{Aggregation and Discretization on Measure.}~Given a measure
$M$, users may perform aggregation operations (such as \texttt{SUM} and
\texttt{AVG} in SQL) over a set of realizations of $M$. In some cases, measures
are processed in the form of a dimension (e.g., use measures for explanations),
which necessitates discretization. It transforms numerical values into several
discrete bins that form a categorical variable.

\noindent\textbf{Filter.}~In this paper, filter is the basic unit of data
operations. Given a multi-dimensional data $D$ and a dimension $X$, a filter
$p_i=\{X=x_i\}$ (e.g., ``Smoking=Yes'') implies an equality assertion to $X$
such that the value of $X$ shall equal $x_i$.

\noindent\textbf{Predicate.}~\revision{The disjunction of filters applied on the
same dimension is a predicate.} Given the dimension $X$, the predicate
$P(x_1,\cdots,x_k)$ is a set containment assertion $\{X=x_1\lor\cdots\lor
X=x_k\}\equiv \{p_1,\cdots,p_k\}$. On a discretized measure, a predicate is an
assertion on ranges. A filter is a special case of a predicate. For clarity, we
represent a general predicate with a capital $P$ and a filter with a lower-case
$p$.

\noindent\textbf{Subspace.}~A subspace is a conjunction of filters on disjointed
dimensions. Given multi-dimensional data $D$, a subspace corresponds to a subset
of rows satisfying the conditions of all filters. \revision{If two subspaces
only differ in one filter, they are regarded as \textit{siblings}. The term
\textit{Context} refers to the variables of two sibling subspaces, where the
\textit{background variables} are the variables with the shared filters and the
\textit{foreground variables} are the variables with the different filters.} In
the following example, we provide a simple instantiation.
\begin{example}
  Consider the preceding dataset in \F~\ref{fig:motivating-example}(a).
  $s=\{\text{Location}=\text{A}\land\text{Lung Cancer}=\text{Severe}\}$
  represents the subspace denoting all patients in ``Location=A'' with severe
  lung cancer. All patients in ``Location=A'' with severe lung cancer and all
  patients in ``Location=B'' with severe lung cancer form a pair of sibling
  subspaces. Here, ``Location'' is the foreground variable and ``Lung Cancer''
  is the background variable.
\end{example}

\noindent \textbf{Selection.}~\revision{We use the following notation to
represent the selection procedure over multi-dimensional data $D$. The subset of
data after the selection operation is defined as $D_{p_i}$, $D_P$, or $D_s$,
where $p_i$ is a filter, $P$ is a predicate, and $s$ is a subspace. We define
$D-D'$ as the rows remaining in $D$ after removing those from $D'$.}

\noindent\textbf{\diff\ and Explanation.}~As illustrated in
\F~\ref{fig:motivating-example}(b), the user would issue a \diff\ to \tool\ for
explanation. We formally define \diff\ as follows.

\begin{definition}[\diff]
    \label{def:diff}
    Given a multi-dimensional data $D$, a user launches aggregate query $agg()$
    on a target measure $M$ under two sibling subspaces $s_1,s_2$. \diff\ is
    defined as $\Delta_{s_1,s_2,M,agg}(D)=agg_M(D_{s_1})-agg_M(D_{s_2})$. For
    brevity, we use $\Delta(D)$ as the shorthand of $\Delta_{s_1,s_2,M,agg}(D)$.
    W.l.o.g., we assume $\Delta$ is always non-negative.
\end{definition}

\begin{example}
  As shown in \F~\ref{fig:motivating-example}(b), we concretize the \diff\
  $\Delta$ with the \texttt{AVG} aggregate on the target ``Lung Cancer'' over
  two sibling subspaces
  $s_1=\{\text{Location}=\text{A}\},s_2=\{\text{Location}=\text{B}\}$, denoting
  the difference in average lung cancer severity in ``A'' and ``B''. 
\end{example}

Indeed, explaining the difference between two aggregate queries is one prevalent
data analysis task. Identifying the cause in data difference constitutes the
basis of many data explanation applications, such as outlier explanation and
data debugging~\cite{glavic2021trends}. In accordance with prior
works~\cite{wu2013scorpion,abuzaid2021diff,glavic2021trends}, we concretize the
problem of XDA by concentrating on the explanation of \textit{data difference}.
The following form is used to provide explanations in response to \diff.

\begin{definition}[Explanation]
  \label{def:explanation}
  Given a \diff, an explanation is represented by the following triplet 
  % \vspace{-5pt}
  \begin{equation}
    \textit{explanation}\coloneqq\tuple{\textit{type}, \textit{predicate}, \textit{responsibility}}
    \vspace{-5pt}
  \end{equation}
\noindent where $\textit{type}\in \{\textit{causal}, \textit{non-causal}\}$
  denotes whether the explanation is causal or non-causal, the
  \textit{predicate} is the content of the explanation, and
  \textit{responsibility}, and a score ranging from 0 to 1 quantifies the extent
  to which the explanation explains the given \diff. 
\end{definition}

\begin{example}
  \F~\ref{fig:motivating-example}(e) lists several explanations to the \diff.
  \F~\ref{fig:motivating-example}(f)-(g) visualize two of them.
  \F~\ref{fig:motivating-example}(f), as a causal explanation, depicts that
  ``Smoking=Yes'' causes the lung cancer severity difference in Location A and B
  with a responsibility of $0.77$.
\end{example}

\parh{Single- vs. Multi-Dimensional Explanation.}~\revision{For conciseness and
clarity, we anticipate that each explanation reflects one aspect contributing to
the outcome when explaining the \diff.
We recommend adopting a single-dimensional explanation in \tool\ due to its
unambiguous causal semantics, although it is feasible to extend an explanation
as multi-dimensional using the Cartesian product. The joint causal semantics of
several variables, however, could be obscure. Furthermore, multiple
single-dimensional explanations (e.g., \F~\ref{fig:motivating-example}(e))
suffice to represent a multi-dimensional case.}

\noindent\textbf{Functional Dependency (FD).}~Functional dependency relations
are common in multi-dimensional data. In a relational database, among the
attributes, there may exist primary keys and foreign keys. Therefore, after
materialization, the resulting multi-dimensional data may have functional
dependencies. A functional dependency between $X$ and $Y$ is represented by
$X\fdarrow Y$. \revision{FD, as a deterministic relation among two variables,
deems a form of reliable knowledge.} This research focuses on one-to-one and
one-to-many FDs. We present a simple exemplary dataset that contains FDs.

\begin{example}
  \label{ex:fd}
  Let \textbf{CityInfo} be a dataset with three attributes (i.e., City, State,
  Country). It has three FDs, namely, $\text{City}\fdarrow \text{State}$,
  $\text{State}\fdarrow \text{Country}$, and $\text{City}\fdarrow
  \text{Country}$.
\end{example}

\noindent\textbf{FD-Induced Graph.}~Given a multi-dimensional data $D$ and its
functional dependencies, the FD-induced Graph $\fdgraph\coloneqq (V,E)$, where
$V\coloneqq \{X_i\mid \forall X_i \in D\}$ and $E\coloneqq \{(X_i, X_j)\mid
\text{if } X_i\fdarrow X_j\}$. We assume $\fdgraph$ to be acyclic. Cycles in
$\fdgraph$ imply redundant attributes; in such cases, we retain only one of
them to ensure acyclicity.

\subsection{Causal Discovery with Latent Variables}
\label{subsec:causal-discovery}

This section presents terminology essential to causal discovery with latent
variables, such as the representation of causal graphs under causal
insufficiency and typical assumptions in causal discovery. 

% Readers can also refer to~\cite{zhang2008completeness,claassen2013learning} for
% more detailed illustration of the following definitions and concepts.

\noindent\textbf{Causal Sufficiency.}~Causal discovery aims to learn the causal
relations from the observational data. Most causal discovery algorithms assume a
sufficient observation of the underlying data generating
process~\cite{spirtes2000causation}. Formally, a set of variables $\bm{X}$ is
said to be causally sufficient if there is no hidden variable $Z\notin\bm{X}$
that is causing more than one variable in $\bm{X}$. \revision{In other words, it
assumes that latent confounders --- the shared causes among two or multiple
variables --- do not exist. However, the process used to acquire real-world data
does not provide such guarantees, thereby often yielding causally insufficient
observations. Hence, the causal discovery procedure is compromised by the
spurious association between two variables sharing a latent confounder}. We
present an example below.

\begin{figure}[!htbp]
\vspace{-10pt}
\centering
\includegraphics[width=0.65\columnwidth]{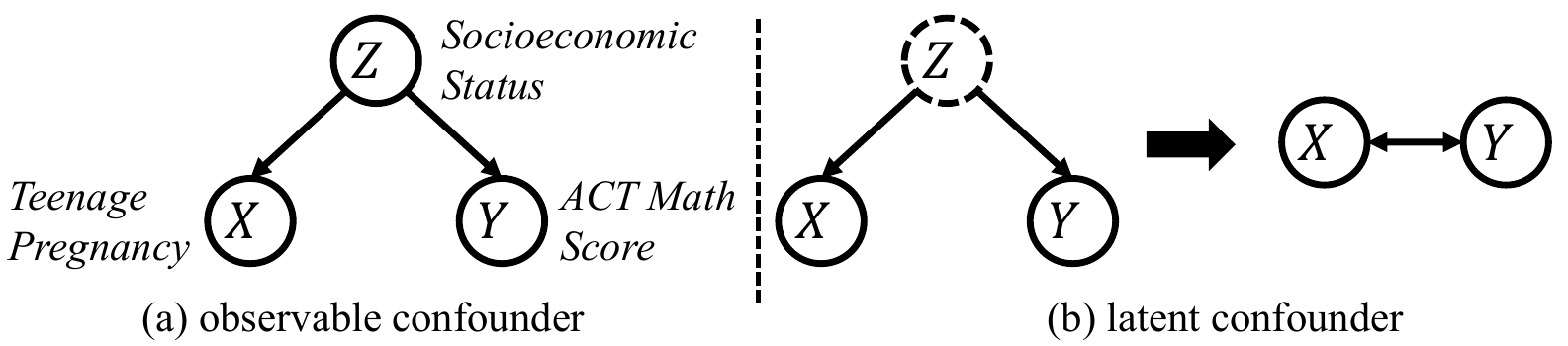}
\vspace{-10pt}
\caption{Examples of observable and latent confounders.}
\label{fig:causal-graph}
\vspace{-10pt}
\end{figure}

\begin{example}
  Consider the \revision{hypothetical causal graph} in
  \F~\ref{fig:causal-graph}(a) where the socioeconomic status ($Z$)
  simultaneously causes teenage pregnancy ($X$) and their ACT math scores ($Y$).
  \revision{The socioeconomic status, however, does not appear in the dataset.
  This absence yields an insufficient observation (the left-side causal graph in
  \F~\ref{fig:causal-graph}(b)), which further results in a spurious
  association~\cite{pearl2009causality} of teenage pregnancy and ACT math scores
  (the bidirected edge in \F~\ref{fig:causal-graph}(b)).}
\end{example}

Hence, popular directed acyclic graphs are not expressive enough to represent
these subtle relations. This necessitates the Maximum Ancestral
Graph~\cite{spirtes2000causation}, which is introduced shortly.

\parh{Notation and Terminology.}~Recall that we assume the dataset is
\textit{i.i.d.}~with potential latent confounders and does not contain selection
bias. Maximal Ancestral Graph (MAG) forms the standard representation of causal
graphs in this setting. We now introduce important concepts of graphical models
and properties of MAG. 

A directed mixed graph $\mathcal{G}$ is a graphical model that contains nodes
$\bm{X}$ and two types of edges, including directed ($\to$) and bidirected
($\leftrightarrow$). There is at most one edge between any two nodes. For each
directed edge $X\to Y$, $X$ is a \textit{parent} of $Y$ and $Y$ is a
\textit{child} of $X$. $X$ and $Y$ are \textit{adjacent} if there is an edge
(either directed or bidirected) between them. A \textit{path} $\mathcal{P}$ is a
sequence of distinct nodes $(X_1, \ldots,X_k)$ where $X_i$ and $X_{i+1}$ are
adjacent in $\mathcal{G}$ for all $1\leq i<k$. A path $\mathcal{P}=(X_1,
\ldots,X_k)$ is directed if $X_i$ is a parent of $X_{i+1}$ for all $1\leq i<k$.
$X$ is an ancestor of $Y$ if there exists a directed path from $X$ to $Y$ and
$Y$ is a descendant of $X$ accordingly. Given a path $(X_1, \ldots,X_k)$, a
non-endpoint node $X_i$ is a collider if there are arrowheads pointing to $X_i$
from both $X_{i-1}$ and $X_{i+1}$. Below, we list all possible cases of a
collider.

\begin{example}
  Given $(X_{i-1},X_i,X_{i+1})$, $X_i$ is a collider if and only if a)
  $X_{i-1}\to X_i\leftarrow X_{i+1}$, or b) $X_{i-1}\leftrightarrow
  X_i\leftarrow X_{i+1}$, or c) $X_{i-1}\to X_i\leftrightarrow X_{i+1}$, or d)
  $X_{i-1}\leftrightarrow X_i\leftrightarrow X_{i+1}$. \revision{In
  \F~\ref{fig:motivating-example}(c), Smoking is a collider of Location and
  Stress since ``Location $\circarrow$ Smoking $\arrowcirc$ Stress'', where
  $\circ$ represents an undetermined edge endpoint.}
\end{example}

\revision{A path $(X,W_1,\cdots,W_k,Y)$ is said to be \textit{blocked} by
$Z\subseteq \bm{X}\setminus\{X,Y\}$ if there exists a node $W_i\in
\{W_1,\cdots,W_k\}$ such that a) $W_i$ is not a collider but a member of $Z$, or
b) $W_i$ is a collider but not an ancestor of any nodes of $Z$.} We now
introduce \textit{m-separation} and MAG.

\begin{definition}[m-separation~\cite{zhang2008completeness}]
  \label{def:mseparation}
  $X,Y$ are m-separated by $Z$ (denoted by $X\sep Y\mid Z$) if
  all paths between $X,Y$ are blocked by $Z$. 
\end{definition}

\begin{example}
  \label{ex:m-sep}
  Consider the causal graph in \F~\ref{fig:motivating-example}(c) where
  ``Smoking'' blocks the only path between ``Location'' and ``Lung Cancer''.
  Hence, ``Smoking'' m-separates ``Location'' and ``Lung Cancer'' (denoted by
  $\text{Lung Cancer} \sep \text{Location}\mid \text{Smoking}$).
\end{example}

\begin{definition}[Maximal Ancestral Graph~\cite{zhang2008completeness}]
  A directed mixed graph is called a MAG if a) it contains no directed cycles or
  almost directed cycles and b) for each pair of non-adjacent nodes, there
  exists a set of nodes that m-separates them. A directed cycle refers to the
  case where $X\to Y\to\cdots\to X$ and an almost directed cycle refers to the
  case where $X\to Y\to\cdots \to Z\leftrightarrow X$.
\end{definition}

\revision{Then, the Global Markov Property (GMP) is developed to provide a
probabilistic interpretation of m-separation.}

\begin{definition}[Global Markov Property~\cite{spirtes2000causation}]
  \begin{equation}
    X \sep Y\mid Z \Rightarrow X \Perp Y\mid Z
    \vspace{-5pt}
  \end{equation}
\end{definition}
As aforementioned, m-separation indicates that all paths between $X$ and $Y$ are
``blocked'' by $Z$. Hence, it is intuitive that, if $X$ and $Y$ are
\textit{m-separated}, their statistical correlation is also ``blocked'' by $Z$.
\revision{The term \textit{conditional independence}
(i.e., $X \Perp Y\mid Z$) depicts this absence of statistical correlation.
Statistically, $X \Perp Y\mid Z$ implies that $P(X,Y\mid Z)=P(X\mid Z)P(Y\mid
Z)$, which can be empirically examined using statistical hypothesis tests
(e.g., $\chi^2$ tests).}
\begin{example}
  Consider the dataset in \F~\ref{fig:motivating-example}(a). According to GMP,
  the m-separation in \Ex~\ref{ex:m-sep} implies that, for the dataset in
  \F~\ref{fig:motivating-example}(a), ``Location'' and ``Lung Cancer'' are
  conditionally independent given ``Smoking,'' in a statistical sense.
\end{example}

With GMP, we can deduce statistical conditional independence in data from
m-separations. Note that only data is available when performing causal
discovery. Hence, we need to invert GMP and establish a connection from data
distribution to the graphical structure. Faithfulness assumption establishes
such connection.
\begin{definition}[Faithfulness~\cite{spirtes2000causation}]
  \begin{equation}
    X \Perp Y\mid Z \Rightarrow X \sep Y\mid Z
  \end{equation}
\end{definition}
\revision{According to faithfulness,} if we observe that two variables are
conditionally independent by a set of variables in data, then they are
\textit{m-separated} by the same set of variables on the causal graph.
Faithfulness and GMP together establish the equivalence between conditional
independence and m-separation and they form the key to causal discovery. In
addition, we define \textit{skeleton} as follows.

\begin{definition}[Skeleton]
  \label{def:skeleton}
  The skeleton $\mathcal{S}$ of a MAG $\mathcal{G}$ is the undirected graph
  obtained by removing all arrowheads from $\mathcal{G}$.
\end{definition}

\parh{Constraint-based Causal Discovery.}~Constraint-based approaches are the
standard solution to causal discovery. With the faithfulness assumption, these
methods exploit the conditional independence derived from data and gradually
establish a MAG $\mathcal{G}$. $\mathcal{G}$ is consistent with all
m-separations entailed by conditional independence. However, there may exist
multiple MAGs that are equally consistent with the m-separations and not
distinguishable, which is called Markov equivalence class, denoted by
$[\mathcal{G}]$. It is worth noting that these feasible MAGs share the same
skeleton while differing in direction on certain edges. These MAGs are
therefore summarized into a compact representation called Partial Ancestral
Graph (PAG) with some undetermined edge endpoints.

\begin{table}[!htbp]
  \vspace{-5pt}
  \centering
  \caption{Four types of edges in PAG. Circle represents undetermined edge
  endpoint (can be either an arrowhead or tail).}
  \vspace{-10pt}
   \resizebox{\linewidth}{!}{
  \begin{tabular}{c|l}
    \hline
    \textbf{Edge} & \textbf{Causal Semantics}\\
    \hline
    $X\rightarrow Y$ & $X$ is a cause of $Y$. \\\hline
    \multirow{1}{*}{$X\leftrightarrow Y$} & neither $X$ nor $Y$ is a cause of each other but they share a latent common cause.\\\hline
    \multirow{1}{*}{$X\circarrow Y$} & 1) $X$ is a cause of $Y$; or 2) neither $X$ nor $Y$ is a cause of each other but they share a latent common cause.\\\hline
    \multirow{2}{*}{$X\circlinecirc Y$} & 1) $X$ may be a cause of $Y$; or 2)
    $Y$ may be a cause of $X$; or 3) neither $X$ nor $Y$ is a cause of each
    other but\\
    &  they share a latent common cause.\\\hline
\end{tabular}
   }
\label{tab:pag-edges}
\vspace{-10pt}
\end{table}

\begin{definition}[Partial Ancestral Graph~\cite{zhang2008completeness}]
  \label{def:pag}
  Let $[\mathcal{G}]$ be a Markov equivalence class of a MAG $\mathcal{G}$. A
  PAG for $[\mathcal{G}]$ is a graph $\mathcal{P}$ with three possible edge
  endpoints (namely, tail, circle and arrowhead; and hence four kinds of edges:
  $\to, \leftrightarrow, \circarrow, \circlinecirc$) such that 1) $\mathcal{P}$
  shares the same adjacencies with $\mathcal{G}$ (and any member of
  $[\mathcal{G}]$), and 2) every non-circle edge endpoint indicates an invariant
  edge endpoint in $[\mathcal{G}]$.
\end{definition}

The second condition in \D~\ref{def:pag} implies that an edge associates a
tail ``$-$'' or arrowhead ``$\to$'' endpoint, if and only if it is invariant in
all $\mathcal{G}\in[\mathcal{G}]$. \T~\ref{tab:pag-edges} lists the semantics of
edges.

\begin{example}
  $\text{Location}~\circarrow~\text{Smoking}$ in
  \F~\ref{fig:motivating-example}~(c) implies that ``Location'' is a cause of
  ``Smoking'' or they share a latent confounder.
\end{example}

\smallskip
\revision{We clarify that the FCI algorithm~\cite{spirtes2000causation}, as a
typical constraint-based approach, consists of two phases. The skeleton of
$[\mathcal{G}]$ is first learned by assuming faithfulness (i.e., the
\texttt{FCI-SL} phase of the FCI algorithm). Then, the undirected edges are
subsequently oriented according to a set of orientation rules (i.e., the
\texttt{FCI-Orient} phase of the FCI algorithm). Finally, the PAG is returned;
see full details of the FCI algorithm in \sm.} 
However, soon we will show that the faithfulness assumption can be violated by
FD relations. In this paper, we focus on establishing a theory and proposing a
solution to tackle this unique challenge that arises in data analysis scenarios.
\revision{That is, our \xlearn\ calibrates the FCI algorithm to correctly
handling FDs (see details in \S~\ref{subsec:cdd}).}

\section{\tool}
\label{sec:xinsight}

\begin{figure}[!htbp]
\centering
\vspace{-10pt}
\includegraphics[width=0.6\columnwidth]{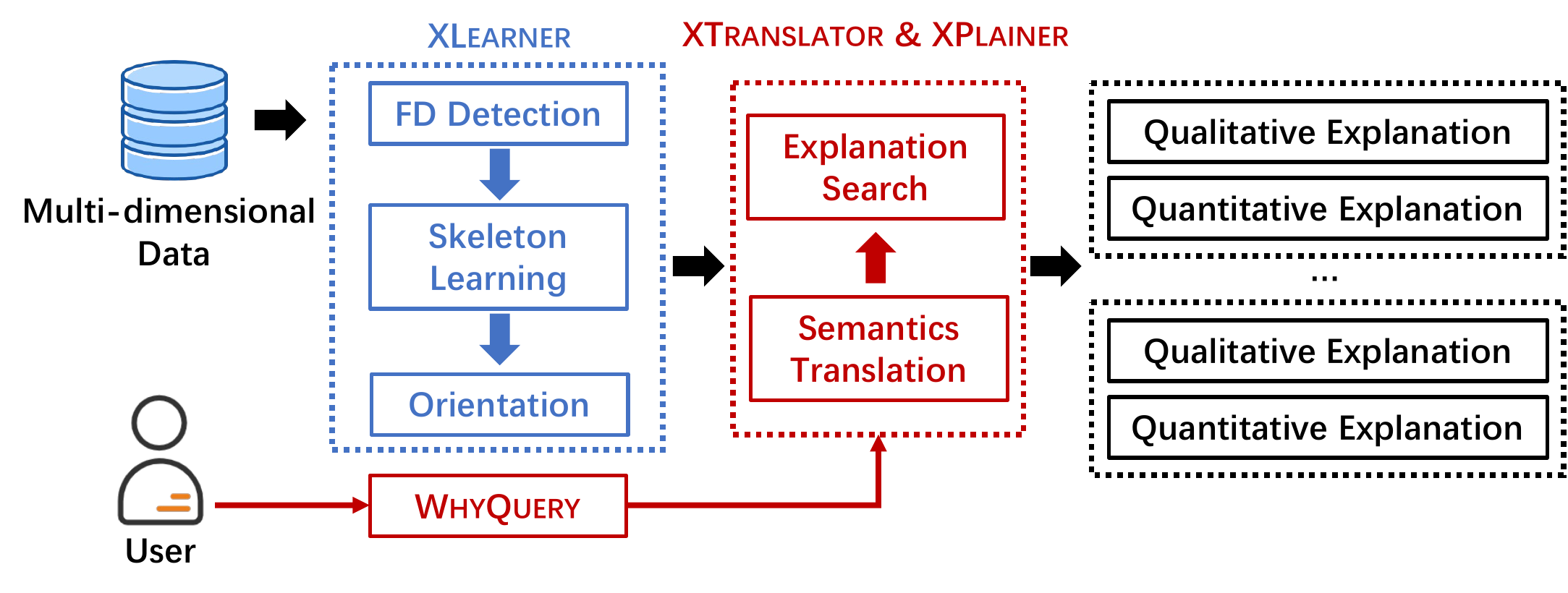}
\vspace{-10pt}
\caption{Workflow of \tool. Offline phase is marked in blue and online phase is marked in red.}
\label{fig:workflow}
\vspace{-10pt}
\end{figure}

\tool\ delivers a unified framework for XDA with three modules. The workflow of
\tool\ is shown in \F~\ref{fig:workflow}. First, given a multi-dimensional data
$D$, \xlearn\ pre-learns a causal graph $\mathcal{G}$ from data in the offline
phase (blue-annotated in \F~\ref{fig:workflow}). Then, in the online phase
(red-annotated in \F~\ref{fig:workflow}), upon receiving a \diff, \xtrans\
identifies variables that have the potential to give either causal or non-causal
explanations based on the causal primitives in $\mathcal{G}$. Finally,
\xplainer\ examines each identified variable with potential and decides the
optimal explanation for the given \diff. After applying \xplainer\ to all
variables with potential, \tool\ yields a set of explanations (with qualitative
sub-explanations and quantitative sub-explanations). By decoupling \tool\ into
an offline phase and an online phase, heavy-weight computations are performed
beforehand, and only light-weight computations are needed in the online phase,
allowing for a rapid response to a user's query. In the following, we elaborate
on the design of each module. Due to page limits, we present proofs and
theoretical discussion in the \sm.

\subsection{\xlearn}
\label{subsec:cdd}

% As noted in \S~\ref{subsec:causal-discovery}, it is hard to guarantee causal
% sufficiency for real-world data. Therefore, \xlearn\ aims to learn a PAG from
% data. Moreover, in the presence of functional dependencies (FD), faithfulness
% can be easily violated~\cite{ding2020reliable}. In the following, we show how FD
% would impose violations of the faithfulness assumption.

\xlearn\ aims to learn a causal graph $\mathcal{G}$ from multi-dimensional data
$D$ in the presence of latent confounders. The primary obstacle is that learning
the skeleton of $\mathcal{G}$ requires the faithfulness assumption (see
\S~\ref{subsec:causal-discovery}), which may be violated by FDs in $D$.
\revision{Below, we show how contradictory causal structures can be induced when
being agnostic to FDs.}

\begin{figure}[!htbp]
\centering
% \vspace{-10pt}
\includegraphics[width=0.6\columnwidth]{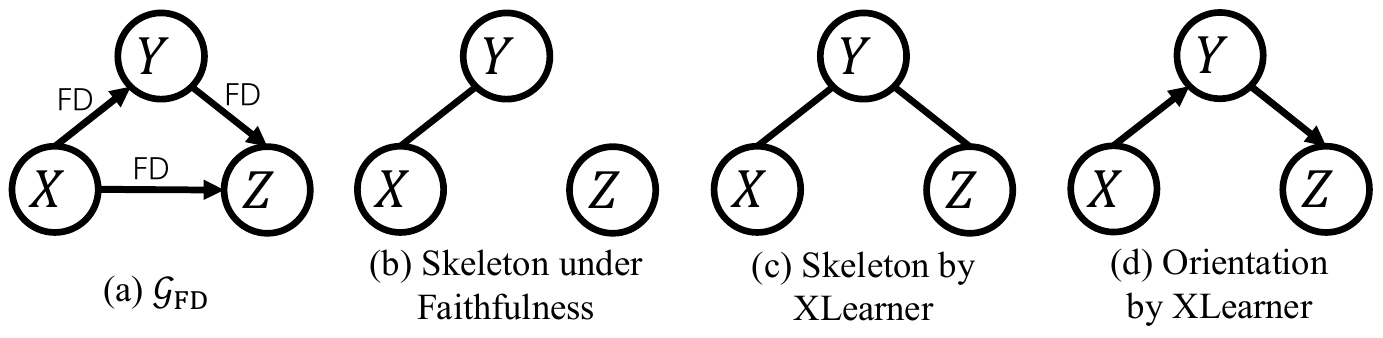}
% \vspace{-10pt}
\caption{Illustration of FD-induced faithfulness violation.}
\label{fig:fd-ex}
% \vspace{-10pt}
\end{figure}

\begin{example}
\label{ex:violation}
We first consider the \textbf{CityInfo} dataset described in \Ex~\ref{ex:fd} and
the corresponding FD-induced graph in \F~\ref{fig:fd-ex}(a), where $X$ denotes
city, $Y$ denotes state, and $Z$ denotes country. By definition of conditional
independence \revision{(i.e., $X \Perp Y\mid Z\iff P(X,Y\mid Z)=P(X\mid
Z)P(Y\mid Z)$)}, we have $Y\Perp Z \mid X$ and $X\Perp Z\mid Y$. \revision{The
definition of faithfulness implies $Y\sep Z \mid X$ and $X\sep Z\mid Y$, given
$Y\Perp Z \mid X$ and $X\Perp Z\mid Y$. $Z$ is non-adjacent to both $X$ and $Y$
according to the m-separation definition (see the induced graph in
\F~\ref{fig:fd-ex}(b)). Consequently, $Z$ is an isolated node in
\F~\ref{fig:fd-ex}(b). We have $Y\sep Z$ and GMP further implies that $Y\Perp
Z$, which contradicts $Y\not\Perp Z$ entailed by $Y\fdarrow Z$.} Indeed, the
skeleton is not consistent with any MAGs that are on the top of it.
\end{example}

\begin{table}[!htbp]
  %\vspace{-5pt}
  \centering
  \caption{Comparing different causal discovery algorithms. \cmark\ denotes
  ``support'' whereas \xmark\ denotes ``no support''.}
  \vspace{-10pt}
   \resizebox{0.8\linewidth}{!}{
  \begin{tabular}{l|c|c|c}
    \hline
    \textbf{Alg.} & \textbf{Orientation} & \textbf{FD-induced Faithfulness Violation}& \textbf{Causal Insufficiency}\\
    \hline
    PC~\cite{spirtes2000causation}  & \cmark & \xmark & \xmark \\\hline
    FCI~\cite{zhang2008completeness}  & \cmark & \xmark & \cmark \\\hline
    REAL~\cite{ding2020reliable}  & \xmark & \cmark & \xmark \\\hline
    \xlearn & \cmark & \cmark & \cmark \\\hline
  \end{tabular}
   }
 \vspace{-5pt}
\label{tab:alg-cmp}
\end{table}

As aforementioned in \S~\ref{sec:preliminary}, the violations of causal
sufficiency and faithfulness (induced by FDs) are common in the data analysis
scenarios. However, they are addressed separately in the literature, as reviewed
in \T~\ref{tab:alg-cmp}. \xlearn\ focuses on addressing both challenges
simultaneously. \F~\ref{fig:fd-ex}(c)-(d) show the skeleton and orientation by
\xlearn, which are compliant with intuition.

\begin{figure}[!htbp]
    % \vspace{-10pt}
\centering
\includegraphics[width=0.6\columnwidth]{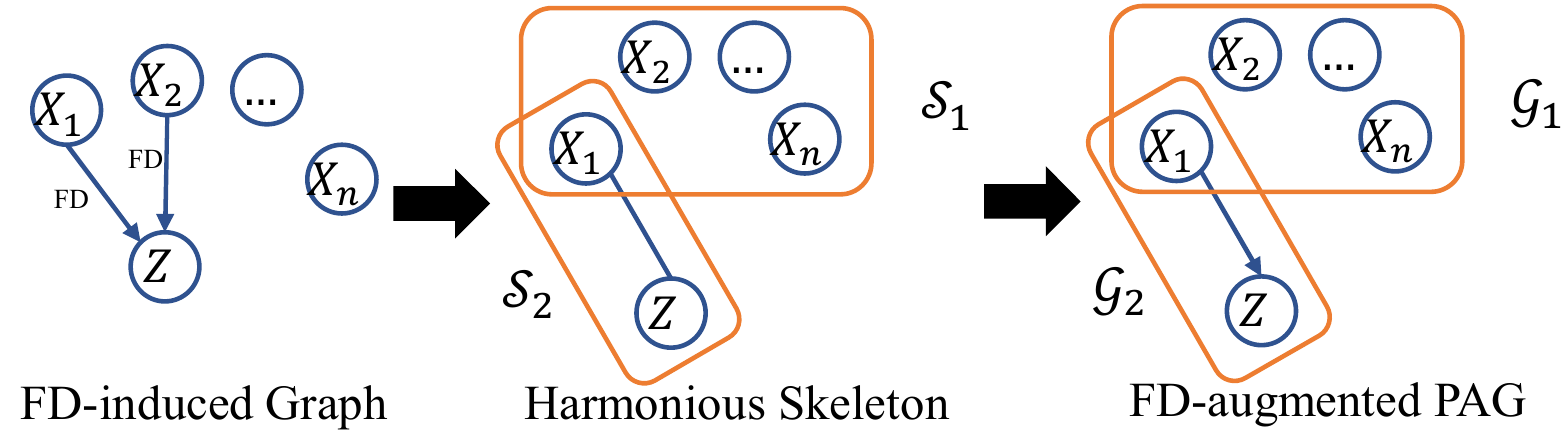}
% \vspace{-10pt}
\caption{Running example of \xlearn.}
\label{fig:single-sink}
% \vspace{-10pt}
\end{figure}

Overall, \xlearn\ tackles the problem in three stages. We outline the workflow
of \xlearn\ in \A~\ref{alg:cdd} and present an example. Then, we elaborate on the
design of \xlearn.
\begin{example}
    Consider the FD-induced graph $\fdgraph$ shown in \F~\ref{fig:single-sink}.
    In the first stage, \xlearn\ uses $\fdgraph$ to identify variables (e.g.,
    $X_1$ and $Z$ in \F~\ref{fig:single-sink}) that may trigger faithfulness
    violations. Then, the skeleton $\mathcal{S}_2$ is built upon a harmonious
    assumption instead of faithfulness over $X_1$ and $Z$. In the second stage,
    the FCI algorithm (skeleton learning and orientation) is only conducted over
    variables that comply with the faithfulness assumption. Hence, the skeleton
    $\mathcal{S}_1$ and the PAG $\mathcal{G}_1$ are identified accordingly. In
    the third stage, we orient $\mathcal{S}_2$ to generate an FD-augmented PAG
    $\mathcal{G}_2$. By concatenating $\mathcal{G}_1$ and $\mathcal{G}_2$, the
    resultant (FD-augmented) PAG $\mathcal{G}$ is obtained.
\end{example}

\begin{algorithm}[!t]
\small
\caption{\xlearn\ procedure.}
\label{alg:cdd}
\KwIn{Multi-dimensional Data $D$, FD-induced graph $\fdgraph$}
\KwOut{FD-augmented PAG $G$}
% $S\leftarrow \slf(D,\fdgraph)$\;
\textcolor{blue}{// stage 1: detect and preclude $\xfd$ (\S~\ref{subsec:skeleton})}\\
$\mathcal{S}_2\leftarrow (V, \emptyset)$\;
Topologically sorting nodes in $\fdgraph$ and record depth as $d(X_i)$\;
\While{$\fdgraph$ has non-root nodes}{
    $X\leftarrow \argmax_{X\in \fdgraph.V} d(X)$\;
    $Y\leftarrow \argmin_{Y\in Pa(\fdgraph, X)} |Y|$\;
    add edge $(X,Y)$ in $\mathcal{S}_2$\;
    remove $X$ and all connected edges from $\fdgraph$\;
}
\textcolor{blue}{// stage 2: standard PAG learning}\\
$\mathcal{S}_1\leftarrow \texttt{FCI-SL}(D,\fdgraph.V)$\;
$\mathcal{G}_1\leftarrow \texttt{FCI-Orient}(S_1)$\;
\textcolor{blue}{// stage 3: orient $\mathcal{S}_1$ and generate $\mathcal{G}$ (\S~\ref{subsec:orientation})}\\
\ForEach{$(X\fdarrow Y)\in\fdgraph.E$}{
    \lIf{$X,Y$ is adjacent in $\mathcal{S}$}{
        orient $X\rightarrow Y$ on $\mathcal{G}^2$
    }
}
generate $\mathcal{G}$ concatenating $\mathcal{G}^1$ and $\mathcal{G}^2$\;
% $G\leftarrow \orientf(S, \fdgraph)$\;
\Return $\mathcal{G}$\;
\end{algorithm}

\smallskip
\parh{{Comparison with FCI.}}~\revision{Comparing with the FCI
algorithm~\cite{spirtes2000causation,zhang2008completeness}, \xlearn\ for the
first time reconciles functional dependency (FD) and the faithfulness assumption
for causally insufficient data within the harmonious skeleton framework. In that
sense, it can learn causal graphs from real-world data adequately. As validated
in \S~\ref{subsec:xlearner-eval}, \xlearn\ learns more accurate causal graphs
than the FCI algorithm. Second, it uses FDs to provide a more complete
orientation to the underlying causal graph. Hence, compared to the FCI
algorithm, it leverages the knowledge from FDs to enforce a more precise causal
graph with less undetermined edges. In sum, we deem that \xlearn\ enhances the
FCI algorithm from the theoretical perspective, and also addresses obstacles in
the real-life adoption of the FCI algorithm.}

\subsubsection{Skeleton Learning with FD (lines 1--9 of \A~\ref{alg:cdd})} 
\label{subsec:skeleton}
Ding et al.~point out that in the presence of FD relations, faithfulness
assumption can be violated thus we can at most obtain a \textit{harmonious
skeleton}~\cite{ding2020reliable}. However, the original theory of
\textit{harmonious skeletons} is established under causal sufficiency. Here, we
further generalize the \textit{harmonious skeleton} for causally insufficient
systems:

% We need to first learn the skeleton (i.e., the undirected graph) while most
% off-the-shelf algorithms are based on faithfulness and GMP. With faithfulness
% and GMP, the learned skeletons are probably correct with respect to the ground
% truth. However, as pointed out in~\cite{ding2020reliable}, under FD-induced
% faithfulness violations, we can at most obtain a \textit{harmonious skeleton}.
% The original definition of \textit{harmonious skeleton} is established under
% causal sufficiency. Here, we further generalize the \textit{harmonious skeleton}
% for causally insufficient systems:

\begin{definition}[Harmonious Skeleton]
    \label{def:harmonious}
    A skeleton $\mathcal{S}$ is said to be harmonious w.r.t.~a joint probability
    distribution $P$ if 1) there exists a MAG $\mathcal{G}$ sharing the same
    adjacencies of $\mathcal{S}$, 2) $P$ satisfies GMP to $\mathcal{G}$, and 3)
    any subgraph of $\mathcal{S}$ does not satisfy the previous two conditions.
\end{definition}

\D~\ref{def:harmonious} entails three properties of $\mathcal{S}$. First, since
there exists a MAG $\mathcal{G}$ on top of the skeleton $\mathcal{S}$, there
exists a set of nodes that m-separates any non-adjacent nodes. Second, if two
nodes (e.g., $X, Y$) are m-separated by $Z$, then $X\Perp Y\mid Z$. These two
conditions imply that $X$ and $Y$ are non-adjacent in $\mathcal{S}$, if and only
if there exists a set of nodes $Z$ such that $X\Perp Y\mid Z$. The last
condition implies the minimality of $\mathcal{S}$, which is commonly
assumed~\cite{peters2017elements}. When two graphs $\mathcal{G},\mathcal{G}'$
are equally compatible with the data, we would prefer the simpler one. We now
show the construction of $\mathcal{S}$, which begins with a basic case and
generalizes to arbitrary structures.

\begin{theorem}
    \label{thm:single-sink}
    Let $Z$ be a sink node (i.e., all edges of $Z$ are oriented to $Z$) in
    $\fdgraph$. $\mathcal{S}=\mathcal{S}_1\cup\mathcal{S}_2$ is a harmonious
    skeleton if 1) $S_1$ is a harmonious skeleton over $\bm{X}\setminus Z$ and
    $\mathcal{S}_2$ contains only one edge $X_i-Z$ where $X_i$ can be any node
    connected to $Z$ in $\fdgraph$.
\end{theorem}

\begin{example}
    \F~\ref{fig:single-sink} presents an example of \Thm~\ref{thm:single-sink}.
    Connecting the sink node $Z$ with one of its parents $X_1$ in the $\fdgraph$
    yields a skeleton $\mathcal{S}_2$. If we can learn a harmonious skeleton
    $\mathcal{S}_1$ on the remaining nodes ($X_1,\cdots,X_n$),
    \Thm~\ref{thm:single-sink} ensures that concatenating
    $\mathcal{S}_1,\mathcal{S}_2$ produces a harmonious skeleton over all
    variables.
\end{example}

According to \Thm~\ref{thm:single-sink}, if $Z$ has more than one parent,
multiple harmonious skeletons exist (note that $Z$ can connect to any one of its
parents). In practice, we connect $Z$ to the parent node with the lowest
cardinality. Given a FD-induced graph, we recursively apply
\Thm~\ref{thm:single-sink} to identify sink nodes and derive the corresponding
harmonious skeleton $\mathcal{S}_2$ until all FDs are properly resolved. Then,
we can apply the standard skeleton learning algorithm over the remaining nodes.
The procedure is shown in lines 1--9 in \A~\ref{alg:cdd}.

% \begin{algorithm}[!htbp]
% \scriptsize
% \caption{Skeleton Learning with FD ($\slf$)}
% \label{alg:slfd}
% \KwIn{Database $D$, FD-induced graph $\fdgraph$}
% \KwOut{Harmonious skeleton $S$}
% $S\leftarrow (V, \emptyset)$\;
% Topologically sorting nodes in $\fdgraph$ and record depth as $d(X_i)$\;
% \While{$\fdgraph$ has non-root nodes}{
%     $X\leftarrow \argmax_{X\in \fdgraph.V} d(X)$\;
%     $Y\leftarrow \argmin_{Y\in Pa(\fdgraph, X)} |Y|$\;
%     add edge $(X,Y)$ in $S$\;
%     remove $X$ and all connected edges from $\fdgraph$\;
% }
% $S\leftarrow S\cup SL(D,\fdgraph.V)$\;
% \Return $S$\;
% \end{algorithm}

\begin{theorem}
    \label{thm:alg}
    The skeleton of \A~\ref{alg:cdd} is harmonious.
\end{theorem}

\A~\ref{alg:cdd} first constructs an empty skeleton $\mathcal{S}$ that shares
the same nodes as $\fdgraph$ (line 2). At line 3, we topologically sort the
$\fdgraph$ nodes (note that $\fdgraph$ is a DAG). In each iteration (lines
5--8), we pick the deepest node and apply \Thm~\ref{thm:single-sink} to connect
$X$ to one of its parents (in $\fdgraph$) $Y$ in the skeleton. We use the parent
node with the lowest cardinality as $Y$ (line 6), as it usually aligns with
human intuition. 
\begin{example}
    Consider the \textbf{CityInfo} dataset in \Ex~\ref{ex:fd}.
    \A~\ref{alg:cdd} identifies the correct skeleton as
    $\text{City}-\text{State}-\text{Country}$ in \F~\ref{fig:fd-ex}(c). 
\end{example}

For root nodes, since there are no FDs and thus the faithfulness assumption
holds, we employ the standard FCI algorithm (lines 10--12) to infer the PAG
$\mathcal{G}_1$. After $\mathcal{S}_2$ being oriented to $\mathcal{G}_2$ (see
\S~\ref{subsec:orientation}), we concatenate them to form $\mathcal{G}$ (line
17). \Thm~\ref{thm:alg} proves that the skeleton of $\mathcal{G}$ is also
harmonious after the concatenation.
% We prove \Thm~\ref{thm:alg} in \sm.

\subsubsection{Orientation (lines 13--16 of \A~\ref{alg:cdd})}
\label{subsec:orientation}
Classical constraint-based causal discovery algorithms decide the direction of
edges based on a set of orientation rules. These rules orient undirected edges on
skeletons (i.e., $\circlinecirc$) based on a set of criteria, including
conditional independence and some graphical structural relations (e.g.,
discriminating path)~\cite{zhang2008completeness}. These rules are 
applied iteratively until no more orientations can be made. 
% As a result, some edges (i.e., $\circlinecirc$)
% may still remain undirected, indicating an unidentifiable causal relations, and
% some edges may be oriented (i.e., $\rightarrow,\leftrightarrow$) or partially
% oriented (i.e., $\circarrow$). 
However, we argue that an FD itself reflects a causal relation to a good extent,
of which the reason is twofold.

\noindent\textbf{ANM Perspective on FD-related Edges.}~We anticipate
incorporating the discrete additive noise model (ANM)~\cite{peters2011causal}
for orienting FD-related edges. The main theory of ANM implies that if an
asymmetric ANM $Y=f(X)+N_Y$ exists from $X$ to $Y$ and $N_Y$ is independent of
$X$, then $X$ causes $Y$. By FD, we note that, if $X\fdarrow Y$ in $\fdgraph$,
an ANM construction from X to Y naturally exist with noise term $N_Y=0$. On the
other hand, an ANM construction from $Y$ to $X$ exists only in very rare cases,
as determined by the identifiability of the discrete ANM (see \Thm~4.6
in~\cite{peters2017elements}). In light of this, we hypothesize that $X\fdarrow
Y$ in $\fdgraph$ implies causation of $X\to Y$. 
% in practice. 
% (see violations of the hypothesis in \sm).

\noindent\textbf{FCI Perspective on FD-related Edges.}~The rules in FCI may be
unreliable due to the faithfulness violations by FDs. However, an FD itself is
generally more reliable, which describes deterministic relations. More
importantly, the directions from the FDs are compatible with the result of the
FCI on the variables excluding FD-related variables. That is, incorporating ANM
would not violate GMP.

% \begin{algorithm}[!htbp]
% \scriptsize
% \caption{Orientation with FD ($\orientf$)}
% \label{alg:orient}
% \KwIn{Skeleton $S$, FD-induced graph $\fdgraph$}
% \KwOut{FD-augmented PAG $G$}
% $G\leftarrow FCI(S)$\;
% \ForEach{$(X\fdarrow Y)\in\fdgraph.E$}{
%     \lIf{$X,Y$ is adjacent in $G$}{
%         orient $X\rightarrow Y$ on $G$
%     }
% }
% \Return $G$\;
% \end{algorithm}

We implement the above hypothesis in our orientation algorithm (lines 13--16 in
\A~\ref{alg:cdd}). We examine, for each FD relation that is also adjacent in
$\mathcal{S}_2$, whether the edge is oriented as $\rightarrow$ (lines 13--15).
We note that, by incorporating ANM, the augmented graph is more informative and
represents an overcomplete graph w.r.t.~the ground-truth MAG's Markov
equivalence class, exhibiting greater precision than causal graphs learned only
by rules.

\subsection{\xtrans}
\label{subsec:csm}

A causal graph does not directly reveal if a variable adequately explains a
\diff, nor does it directly reflect if the variable features a causal or
non-causal explanation. Bridging this gap requires a translation from causal
primitives to XDA semantics. To illustrate, we start with a \diff\ under
\texttt{AVG}. We then show how to generalize the main result into \texttt{SUM}
and other aggregates.

\parh{Principle of Explainability.} Given a \diff\ $\Delta$ where
$agg=\texttt{AVG}$, a variable $X$ is said to have \textit{No Explainability} if
$X\Perp M\mid F\cup \boldsymbol{B}$, where $M$ is the target measure, $F$ is the
foreground variable, and $\boldsymbol{B}$ are background variable(s). In the
subsequent discussion, we omit $\boldsymbol{B}$ for the ease of presentation
without loss of generality.

A \diff\ in XDA requires us to observe the difference between aggregates on $M$
within two subspaces. The conditional independence of $X\Perp M\mid F$ implies
that $\mathbb{E}(M\mid F, X)=\mathbb{E}(M\mid F)$. Hence,
$\Delta(D)=\Delta(D_{X=x})$ in the large sample limit for all feasible filters
in $X$. If $X$ is \textit{conditionally independent} of $M$ given $F$, $X$ is
simply impossible to offer explanations to the \diff.
Thus, this principle imposes a restriction on possible variables that have
the potential to provide explanations. In particular, we derive the following
restriction.

\begin{proposition}
    \label{prop:mseparation}
    If $X$ has explainability, $M,X$ are not m-separated by $F$ in the causal
    graph $G$.
\end{proposition}

\Prop~\ref{prop:mseparation} illustrates the chance of pruning variables for
which it is impossible to provide explanations. \T~\ref{tab:pag-paths} further
depicts the translation from causal primitives to XDA semantics. In \xtrans, a
variable $X$ is first confirmed to have explainability if $X,M$ are not
m-separated by $F$ in $G$ (1st row in \T~\ref{tab:pag-paths}). In addition,
\xtrans\ also categorizes whether $X$ is causal or non-causal according to
\T~\ref{tab:pag-paths}. Overall, $X$ provides a causal explanation if it is
explainable and a cause (\ding{192} and \ding{193} in \T~\ref{tab:pag-paths}) or
a possible cause (\ding{194} and \ding{195} rows in \T~\ref{tab:pag-paths}) of
$M$. We show how the causal graph identified by \xlearn\ is translated.

% This is intuitive, as an explanation should serve as a causal factor for the
% target, but not necessarily as the direct cause. 

\begin{example}
    Given the dataset in \F~\ref{fig:motivating-example}(a), \xlearn\ identifies
    the corresponding causal graph in \F~\ref{fig:motivating-example}(c). With
    the \diff\ in \F~\ref{fig:motivating-example}(b), \xtrans\ translates the
    causal graph into the XDA semantics in \F~\ref{fig:motivating-example}(d).
    ``Smoking'' and ``Stress Level'' can be used to \textit{causally} explain
    ``Lung Cancer''. And, other variables (e.g., ``Surgery'') are deemed
    non-causal explanations (last row in \T~\ref{tab:pag-paths}).
\end{example}

% For example, ``Smoking'' can be used to
% explain ``Lung Cancer'' (direct cause), and ``Stress Level'' can also be used to
% explain ``Lung Cancer'' (indirect cause). All other cases are conservatively
% treated as giving non-causal explanations (last row in \T~\ref{tab:pag-paths}).
% It is possible to derive a finer-grained explanation semantics within the
% non-causal explanation category, such as differentiating the cases where
% $M\rightarrow X$ or $M\leftrightarrow X$. We leave this as a future work.

\begin{table}[!htbp]
    \vspace{-10pt}
      \centering
  \caption{Translating causal primitives to XDA semantics.}
  \vspace{-10pt}
   \resizebox{0.5\linewidth}{!}{
  \begin{tabular}{c|c|c|c}
    \hline
    \textbf{Rule} &\textbf{Path} &  \textbf{Causal Primitive} & \textbf{XDA Semantics}\\
    \hline
    \ding{192} & $X\rightarrow F\rightarrow M, \cdots$& m-separated & no explainability \\\hline
    \ding{193} & $X\rightarrow M$& parent & causal explanation \\\hline
    \ding{194} & $X\rightarrow\cdots\rightarrow M$& ancestor & causal explanation \\\hline
    \ding{195} & $X\circarrow M$& almost parent & causal explanation \\\hline
    \ding{196} & $X\circarrow\cdots \circarrow M$& almost ancestor & causal explanation \\\hline
    \ding{197} & others & N/A & non-causal explanation\\\hline
\end{tabular}
   }
\label{tab:pag-paths}
\vspace{-10pt}
\end{table}

\parh{Extension to \texttt{SUM}.}~The above formulation over
\textit{explainability} is established on \texttt{AVG} aggregates. In the
following, we discuss the implications of our formulation on \texttt{SUM}
aggregates. If $X$ has no explainability, $X\Perp M\mid F$. When we enforce
$X=x$, $\Delta(D_{X=x})$ can merely be affected by the number of rows where
$X=x$ in two sibling subspaces (namely a \texttt{COUNT}-based explanation)
instead of a causal relation between $X$ and $M$ (see detailed formulation in
\sm). This may be valid for explanations; nevertheless, it is typically
inconsistent with the common intuition of data analysis and may not align user
expectations regarding explanations (i.e., a variable explains the target). Such
\texttt{COUNT}-based explanation is unconventional and is thus less of a
concern. 

\parh{Semantics Consistency.}~\revision{Following the above discussion, we
clarify that a variable may play different roles in various aggregates. However,
in our current design, \xtrans\ focuses primarily on variables with strong
connections to $M$, which are more likely to provide desirable explanations.
Therefore, the semantics of a variable are consistent across different
aggregates. As clarified in \textbf{Principle of Explainability} above, it is
appropriate for pruning uninformative variables from general aggregates, and we
do not observe notable issues in practice. We leave designing more comprehensive
translation rules for future research.}

% With above considerations in mind, we follow the seminal research on
% explanation~\cite{keil2006explanation} from the psychological community and
% outline eight useful graphical structures that provide meaningful causal
% explanations in \T~\ref{tab:pag-paths}. We note that in decision-making, a
% direct \textit{explanation} is the most desirable since it triggers
% counterfactual thoughts to explain why the difference exists. In addition to
% \textit{explanation} which is a clear cause of $A_{agg}$, we also consider the
% cases of \textit{coherence} and \textit{relevance}. \fm{Consider the example
% stated in~\cite{keil2006explanation} again.} Initial symptoms of a bacterial
% infection may include an elevation in white blood cell count and fever, followed
% by anemia and headache. When a user is astonished by the high blood cell
% elevation population in a given year, including a high proportion of anemia and
% headache symptoms in explanations not only provides a more holistic view for
% data understanding, but also helps confirm the unexpected query outcome.
% Furthermore, by incorporating PAG to depict causal relations, users are made
% aware of unseen confounders. In the above example, when bacterial infections may
% not be observed in the data, a bidirectional edge $\leftrightarrow$ between
% blood cell elevation and fever enables users to \fm{connect with data more
% deeply} by incorporating their domain expertise.

\subsection{\xplainer}
\label{subsec:xplainer}

\xlearn\ and \xtrans\ together provide a coarse-grained, variable-level qualitative
explanation to a \diff. For instance, ``Smoking''  is a causal explanation for
the differences in severity of ``Lung Cancer'' in Locations A and B.
% In real-world scenarios, a plausible explanation shall also ship
% quantitative explanation. Recalling the example mentioned in
% \S~\ref{sec:introduction}, users would anticipate to know that ``Smoking=Yes''
% to what extent explains high lung cancer severity, but \xtrans\ only tells that
% ``Smoking'' is a cause of ``Lung Cancer''. 
To go one step further, \xplainer\ provides predicate-level quantitative
explanations to answer \diff\ (e.g., ``Smoking=Yes'' explains the difference
with the responsibility of $0.77$ in \F~\ref{fig:motivating-example}(f)).
\xplainer\ is on the basis of a well-establish framework, DB
causality~\cite{meliou2010causality} (an extension of actual causality).
\revision{To ease reading, below we first provide a recap of the notations
defined in \S~\ref{subsec:prel-data}. We then rewrite the formulation of DB
causality in the context of \tool\ in \S~\ref{subsubsec:adaption}.}

\parh{Recap of Notations.}~\revision{We refer to a dataset as $D$, a filter as a
lowercase $p$, and a set of filters as an uppercase $P$. The subset of $D$
satisfying $p$ (or $P$) is represented by $D_p$ (or $D_P$, respectively). We use
$D-D_P$ as the complement of $D_P$ in $D$. By default, $\Delta(D)$ represents
the \diff\ over the dataset $D$. Likewise, for arbitrary $D'\subseteq D$,
$\Delta(D')$ represents the difference between the aggregated values of two
sibling subspaces inside $D'$.}

\begin{definition}[DB Causality~\cite{meliou2010causality}]
    \label{def:db-causality}
    Given a multi-dimensional data $D$ and \diff\ $\Delta$, let $t$ be a tuple
    in $D$. $t$ is called a \textit{counterfactual cause} to $\Delta$, if
    $\Delta(D)>\epsilon$ and $\Delta(D-\{t\})\leq \epsilon$, where $\epsilon$ is
    a user-defined threshold. $t$ is called an \textit{actual cause} to
    $\Delta$, if there exists a contingency $\Gamma\subseteq D$ such that $t$ is
    a \textit{counterfactual cause} for $D-\Gamma$ (i.e.,
    $\Delta(D-\Gamma-\{t\})\leq \epsilon<\Delta(D-\Gamma)$).
\end{definition}

\begin{definition}[DB Responsibility~\cite{meliou2010causality}]
    \label{def:db-responsibility}
    Suppose $P$ is an actual cause to \diff\ $\Delta$ and $\Gamma$ ranges over
    all valid contingencies for $P$. The responsibility of $P$ is defined as
    $\rho_P=\frac{1}{1+\min_\Gamma |\Gamma|}$, where $|\Gamma|$ denotes the
    number of tuples in the contingency.
\end{definition}

DB causality is appealing as it offers both a normalized measure (responsibility
$\in (0,1]$) and a contingency. First, when the responsibility is close to 1, it
implies that the tuple is more accountable for the outcome, and when it hits 1,
it is totally responsible. Second, the minimal contingency reflects the
additional influential factors that, together with the tuple, are fully
responsible for the outcome. The two elements form a quantitative explanation
and it is useful for users to understand why the difference exists.

% DB causality provides clear semantics to measure the responsibility of an
% explanation. Here, responsibility defined as a absolute value ranging from 0 to
% 1. When the responsibility of an explanation is 1, it indicates that the
% explanation is \textit{fully responsible}. In the cases of responsibility $<1$,
% it further supplies contingency and counterfactual semantics, allowing users to
% investigate other influential factors. We deem it as useful because it provides
% a clear and meaningful quantitative score over explanations, while many
% heuristically designed scoring methods can only be used to compare relative
% usefulness of explanations. 

% Various characteristics in database scenarios make it infeasible to directly use
% the method of actual causality. In DB community, there already exists an
% adaptation of actual causality called DB causality~\cite{meliou2010causality}.
% We rewrite its original definition under the context of \diff\ as follows.

\subsubsection{Adaption}
\label{subsubsec:adaption}

DB causality was originally designed for data provenance. As pointed out
in~\cite{meliou2014causality}, tuple-level explanations are usually too
fine-grained for data analysis scenarios. An individual tuple usually has too
little effect on the highly aggregated outcome of a large dataset. Recalling the
example in \F~\ref{fig:motivating-example}, users would expect to know that
``Smoking=Yes'' causes high ``Lung Cancer'' severity rather than an individual
patient being the cause of the high ``Lung Cancer'' severity. This necessitates
predicate-level explanations which are easier to understand and frequently used
in data analysis scenarios, and by many data explanation
tools~\cite{wu2013scorpion,abuzaid2021diff}. Motivated by this, we make three
adaptions over DB causality (namely, \textsc{W-Causality},
\textsc{W-Responsibility} and conciseness) to support XDA. We formulate
\textsc{W-Causality} as follows.

\begin{definition}[W-Causality]
    \label{def:w-causality}
    Given a multi-dimensional data $D$, an attribute of interest $X$ and \diff\
    $\Delta$, let $P\subseteq \bigcup p_i$ be a predicate in $D$, where $\bigcup
    p_i$ denotes the set of all possible filters on $X$. $P$ is called a
    \textit{counterfactual cause} of $\Delta$, if $\Delta(D)>\epsilon$ and
    $\Delta(D-D_P)\leq \epsilon$, where $\epsilon$ is a user-defined threshold.
    $P$ is deemed to be an \textit{actual cause} of $\Delta$, if there is a
    contingency $\Gamma\subseteq \bigcup p_i$ such that $P$ is a
    \textit{counterfactual cause} for $D-D_\Gamma$ (i.e.,
    $\Delta(D-D_\Gamma-D_P)\leq \epsilon<\Delta(D-D_\Gamma)$), where
    $P\cap\Gamma=\emptyset$.
\end{definition}

\noindent\textbf{From Tuples to Predicates.}~\D~\ref{def:w-causality} transforms
the tuple and contingency into two predicates. This way, explanations as well as
contingencies constitute a form of intervention over the multi-dimensional data.
When a contingency $\Gamma$ is applied, it indicates that, if the events related
to $\Gamma$ do not happen, then the events related to $P$ are fully responsible
for $\Delta$. This adaption in turn entails another adaption to the
responsibility for predicate-level explanations.

\begin{definition}[W-Responsibility]
    \label{def:w-responsibility}
    Suppose $P$ is an actual cause to \diff\ $\Delta$ and $\Gamma$ range over
    all valid contingencies for $P$. The responsibility of $P$ is defined as
    $\rho_P=\frac{1}{1+\min_\Gamma |\Gamma|_W}$, where $|\Gamma|_W$ is defined
    as $\max(\frac{\Delta(D-D_P)-\Delta(D-D_P-D_\Gamma)}{\Delta(D)},0)$. We let
    $\rho_P=0$ if $P$ is not an actual cause.
\end{definition}

\noindent\textbf{\textsc{W-Responsibility}.}~Instead of using the number of rows
in $D_\Gamma$ as $|\Gamma|_W$, \D~\ref{def:w-responsibility} employs the
truncated difference in $\Gamma$ over $\Delta$ to measure the importance of $P$.
In particular, $\Delta(D-D_P)-\Delta(D-D_P-D_\Gamma)$ can be deemed as
\textit{first-order finite backward difference} to the function $\Delta(\cdot)$
at the point of $D-D_P$ and $\Gamma$ is the step size. This supplies a simple
and intuitive way to understand to what extent $\Gamma$ plays an important role
in reducing the difference. The large difference imposed by $\Gamma$ implies a
low importance of explanation $P$ to $\Delta$; because the reduction in $\Delta$
is primarily caused by $\Gamma$ instead of $P$ itself. Therefore, the
responsibility of $P$ is measured by a valid contingency $\Gamma^*$ (such that
$\Delta(D-D_P-D_\Gamma)\leq\epsilon$ and $\Delta(D-D_\Gamma)>\epsilon$) with
minimal difference on $\Delta$. 
% In summary, \D~\ref{def:w-causality}
% specifies the property of a valid explanation, whereas
% \D~\ref{def:w-responsibility} quantifies the degree of a valid explanation.

\noindent\textbf{Conciseness.}~Using responsibility as the sole
criterion is not sufficient in practical data analysis
scenarios~\cite{glavic2021trends}. Typically, a \textit{concise} explanation is
preferable. Therefore, given an attribute of interest $X$, we formulate the
optimal explanation of $X$ as follows.
\begin{equation}
    \label{eq:optimization}
    \argmax_{P\subseteq \bigcup p_i} \rho_P - \sigma |P|
\end{equation}
where $\bigcup p_i$ is the set of all possible filters in $X$, $|P|$ is the
number of filters in $P$, and $\sigma |P|$ ($\sigma>0$) forms a conciseness
regularization. In practice, we would prefer $\sigma=\sfrac{1}{m}$ such that 
when all filters are picked, the score is zero.
% \begin{figure}[!htbp]
% \centering
% \includegraphics[width=\columnwidth]{fig/xplainer.pdf}
% \vspace{-10pt}
% \caption{Different search solutions in \xplainer. Dots denote filters in the
% dimension; dots with colored background denote true match; and dots in the
% middle circle denote the explanation found in reality.}
% \label{fig:search}
% \vspace{-15pt}
% \end{figure}

\begin{table}[!htbp]
    \vspace{-7pt}
  \centering
  \caption{Different search solutions in \xplainer. FP is false positive and FN 
  is false negative.}
  \vspace{-10pt}
   \resizebox{0.5\linewidth}{!}{
  \begin{tabular}{l|c|c}
    \hline
    \textbf{Solution} & \textbf{Complexity} & \textbf{Optimality}\\
    \hline
    Brute-force Search & $O(2^m)$ & Optimal \\\hline
    Approx. Search (\texttt{SUM})  & $O(m\log m)$ & Moderated FP; Negligible FN \\\hline
    Approx. Search (\texttt{AVG})  & $O(m^2)$ & Moderated FP\&FN \\\hline
  \end{tabular}
   }
   \vspace{-10pt}
   \label{tab:search}
\end{table}

\subsubsection{Optimization}

As pointed out in~\cite{bertossi2020score,bertossi2020causality}, computing
responsibility (i.e., $\rho_P$) is intractable. Furthermore, solving the
optimization problem in \E~\ref{eq:optimization} is itself difficult given $2^m$
search space ($m$ is the number of filters in $X$). We characterize the
performance of different solutions in \T~\ref{tab:search}. First, the
brute-force search is the most accurate and general method for arbitrary
aggregates, despite being very slow. The explanation discovered by brute-force
search is exactly the optimal explanation. In this paper, we design two
approximate solutions for \texttt{SUM} and \texttt{AVG}, respectively. In
particular, we first show the existence of a linearithmic approximated solution
when the aggregation is \texttt{SUM}. This solution also has negligible false
negatives, as theoretically guaranteed by a lemma on the completeness.
Furthermore, we present a heuristics-based solution for \texttt{AVG} 
%(and
%potentially other aggregates) 
with quadratic complexity. Moreover, this solution should be applicable for
other aggregate functions with a mild downgrade in optimality. Our evaluation
(\S~\ref{subsec:rq3}) shows that both approximations are tight and efficient in
comparison to brute-force search.

\noindent\textbf{Optimization for \texttt{SUM}.}~Given the additive property of
\texttt{SUM} (i.e., $\Delta(D_{P_1}+D_{P_2})= \Delta(D_{P_1}) +
\Delta(D_{P_2})$), we obtain the following proposition to prune the
search space.

\begin{proposition}
    \label{prop:positive}
    If $P^*$ is the optimal explanation of \E~\ref{eq:optimization}, $\forall
    p\in P^*, \Delta(D_{p})>0$.
\end{proposition}

% We provide the proof of \Prop~\ref{prop:positive} in \sm. 

According to \Prop~\ref{prop:positive}, the search algorithm can omit filters
with a non-positive $\Delta_i$ (i.e., $\Delta(D_{p_i})$). Recall that
\E~\ref{eq:optimization} seeks the optimal explanation. When the aggregate
function is \texttt{SUM}, we only need to focus on filters with a reasonably
high $\Delta_i$ without losing optimality and we define such filters as
\textit{canonical filters}.

\begin{definition}[Canonical Filter and Predicate]
    Without loss of generality (w.l.o.g.), given a \diff\ $\Delta$ and an
    attribute of interest $X$, let filters $\{p_1,\cdots,p_m\}$ of $X$ be
    ordered by $\Delta_i$ (i.e., $\Delta(D_{p_i})$) such that $\Delta_1\geq
    \cdots\geq \Delta_m$. We let $p_1,\cdots,p_j$ be canonical filters if
    \begin{equation}
        \label{eq:canonical}
        \Delta(D)-\sum_{i=1}^{j} \Delta_i\leq \epsilon<\Delta(D)-\sum_{i=1}^{j-1} \Delta_i
    \end{equation}
    % Accordingly, 
    $P^C=\{p_1,\cdots,p_j\}$ is called a canonical predicate and 
    $\tau=\sum_{i=1}^{j} \Delta_i$.
\end{definition}

With canonical filters and a corresponding canonical predicate $P^C$, we observe
that $P^C$ manifests good properties. First, $P^C$ is the minimal counterfactual
cause entailed by \E~\ref{eq:canonical}. Our construction of canonical
predicates guarantees completeness.
\begin{proposition}[Completeness]
    \label{prop:completeness}
    For \texttt{SUM}, given a \diff\ $\Delta$, an attribute of interest $X$ and
    corresponding canonical predicate $P^C$, there exists an optimal explanation
    $P^*\subseteq P^C$.
\end{proposition}

The completeness proposition (\Prop~\ref{prop:completeness}) allows us to only
focus on canonical filters when searching for the optimal explanation without
loss of optimality. More importantly, the canonical predicate also allows us to
efficiently identify actual causes and the corresponding valid contingencies.

\begin{theorem}
    \label{thm:complement-set-contingency}
    For \texttt{SUM}, given a \diff\ $\Delta$, an attribute of interest $X$ and
    corresponding canonical predicate $P^C$, $\forall P\subset P^C$, $P$ is an
    actual cause and $\overline{P}=P^C - P$ is a valid contingency.
\end{theorem}

% We provide the proof of \Thm~\ref{thm:complement-set-contingency} in \sm. 

The advantages of \Thm~\ref{thm:complement-set-contingency} are twofold. First,
we can directly confirm valid explanations without exhaustive enumerations.
Second, by the property of $\overline{P}$, we bound $P$'s responsibility
($\rho_P$).

\begin{theorem}
    \label{thm:approx}
    For \texttt{SUM}, given a \diff\ $\Delta$, an attribute of interest $X$ and
    corresponding canonical predicate $P^C$, the \textsc{W-Responsibility}
    $\rho_P$ of $P\subset P^C$ satisfies
    \begin{equation}
        \frac{1}{1+\frac{\tau-\Delta(D_P)}{\Delta(D)}}\leq \rho_P\leq \frac{1}{2-\frac{\Delta(D_P)+\epsilon}{\Delta(D)}}
    \end{equation}
    When $\Delta(D_P) \ll \tau $ and $0<\tau \leq \Delta(D)$,
    $\frac{1}{1+\tau-\Delta(D_P) }\approx\frac{1+\tau +\Delta(D_P)
    }{(1+\tau)^2}$ and the corresponding approximation error rate $E< 0.25$.
\end{theorem}

\Thm~\ref{thm:approx} provides a way to efficiently approximate responsibility
with theoretical guarantees. In that sense, we can compute responsibility
immediately and alleviate searching the minimal contingency. Let
$\hat{\rho_P}=\frac{1+\tau +\Delta(D_P) }{(1+\tau )^2}$, we can rewrite the
objective function in the following form.
\begin{equation}
    \label{eq:approx-optimization}
    % \begin{aligned}
    %     \hat{\rho_P} - \sigma |P|&=\frac{1}{1+\frac{\tau}{\Delta(D)} }+\frac{1}{(1+\frac{\tau}{\Delta(D)})^2}\times \frac{\Delta(D_P)}{\Delta(D)}  - \sigma|P|\\
    %     & = \frac{1}{1+\frac{\tau}{\Delta(D)} }+\frac{1}{(1+\frac{\tau}{\Delta(D)})^2}\times \sum_{p_i\in P} (\frac{\Delta_i}{\Delta(D)}-\frac{\sigma}{(1+\frac{\tau}{\Delta(D)})^2})\\
    %     & = C_1 + C_2\times \sum_{p_i\in P} (\Delta_i -C_3)
    % \end{aligned}
    \hat{\rho_P} - \sigma |P|= C_1 + C_2\times \sum_{p_i\in P} (\Delta_i -C_3)
\end{equation}
Here, $C_1,C_2,C_3$ are constants.
% Given that $\tau,\sigma,\Delta(D)$ are constants during optimization, we
% simplify the optimization problem as follows.
% \begin{equation}
%     \label{eq:optimization-simplified}
%     \argmax_{P\subseteq P^C} \sum_{p_i\in P} (\frac{\Delta_i}{\Delta(D)} -\frac{\sigma}{(1+\frac{\tau}{\Delta(D)})^2})
% \end{equation}
Then, the optimal explanation to \E~\ref{eq:approx-optimization} is
straightforward:
% compute $\Delta_i$ for each $p_i\in P^C$ and the optimal explanation is
\begin{equation}
    P^*=\{p_i\mid \Delta_i>C_3\}
\end{equation}
where $C_3=\frac{\sigma \Delta(D)}{(1+\frac{\tau}{\Delta(D)})^2}$. The
complexity is $\mathcal{O}(m\log (m))$ (primarily in sorting filters for
generating canonical predicates).

\noindent\textbf{Optimization for \texttt{AVG}.}~ In terms of \texttt{AVG}, it
is generally much more challenging due to the absence of the additive
characteristics on $\Delta(D_P) $. Therefore, the majority of the preceding
propositions are not applicable. Having said that, we find the causal graph 
gives considerable opportunities to prune unnecessary computations.
% some properties of
% homogeneous \texttt{AVG} still gives considerable optimization opportunities. 

\begin{definition}[Homogeneous Sibling Subspace]
    Given sibling subspaces $s_1,s_2$ (with foreground variable $F$ and
    background variables $\boldsymbol{B}$), an attribute $X$ and the causal
    graph $G$, $s_1,s_2$ are homogeneous on $X$ if $X,F$ are m-separated given
    $\boldsymbol{B}$ on $G$.
\end{definition}

\begin{proposition}
    \label{prop:avg-inequality}
    For a homogeneous \texttt{AVG}, given a \diff\ $\Delta$, an attribute of
    interest $X$, a predicate $P\subseteq \bigcup p_i$ and a filter $p\in P$, if
    $\Delta(D_p)>\Delta(D_P)$, then $\Delta(D_{P}-D_p)<\Delta(D_P)$.
\end{proposition}

To practically address the search problem of \texttt{AVG}
(\E~\ref{eq:optimization}), we rely on greedy-based heuristics with a pruning
strategy enabled by \Prop~\ref{prop:avg-inequality}. The algorithm is outlined
in \A~\ref{alg:xplainer-avg}.

\setlength{\textfloatsep}{5pt}% Remove \textfloatsep
\begin{algorithm}[!t]
\small
\caption{\xplainer\ For \texttt{AVG}}
\label{alg:xplainer-avg}
\KwIn{\diff\ $\Delta$, threshold $\epsilon$, consiseness parameter $\sigma$}
\KwOut{(near) optimal explanation $P^*$}
$P^C \leftarrow \emptyset$\;
\ForEach{$r=1,\cdots,\min(m,\frac{1}{\sigma})$}{
    \lIf{$\Delta(D-D_{P^C})\leq \epsilon$}{
        break
    }
    \Else{
        $\overline{P}\leftarrow \{p_1,\cdots,p_m\}-P^C$\;
        \uIf{homogeneous}{
            $S\leftarrow \{p_i\mid p_i\in \overline{P}, \Delta_i>\Delta(D-D_{P^C})\}$\;
            $p^*\leftarrow \argmin_{p\in S} \Delta(D-D_{P^C}-D_{p})$;
        }
        \Else{
            $p^*\leftarrow \argmin_{p\in \overline{P}} \Delta(D-D_{P^C}-D_{p})$;
        }
        % $p^*\leftarrow \argmin_{p\in S} \Delta(D-D_{P^C}-D_{p})$\;        
        $P^C\leftarrow P^C\cup \{p^*\}$\;
    }
}
\lIf{$\Delta(D-D_{P^C})> \epsilon$}{\Return{$\perp$}}
\ForEach{$k\in 1,\cdots,|P^C|$}{
    $P_k \leftarrow$ top-k filters of $P^C$\;
    $\Gamma_k \leftarrow P^C-P_k$\;
    compute $\hat{\rho_{P_k}}$ with $\Gamma_k$.
}
\Return $\argmax_k \hat{\rho_{P_k}}-\sigma |P_k|$\;
\end{algorithm}

The high-level idea behind \A~\ref{alg:xplainer-avg} is similar to the one for
\texttt{SUM}, which attempts to construct a canonical predicate $P^C$ such that
$P^C$ forms a counterfactual cause, each subset $P\subset P^C$ of the canonical
predicate constitutes an actual cause, and the complement set $P^C-P$ is a valid
contingency. Unlike \texttt{SUM}, however, \A~\ref{alg:xplainer-avg} does not
ensure the optimality of the resulting explanation, primarily due to the
incompleteness of the canonical predicate (\Prop~\ref{prop:completeness}) under
\texttt{AVG}. Recall that $\rho_P$ ranges from 0 to 1 in
\E~\ref{eq:optimization}. The optimal explanation contains at most
$\sfrac{1}{\sigma}$ filters (otherwise, $\rho_P-\sigma|P|<0$). Hence, the
canonical predicate $P^C$ shall contain at most $\sfrac{1}{\sigma}$ or $m$
(i.e., the number of filters in the attribute). 

\A~\ref{alg:xplainer-avg} employs a greedy strategy to construct $P^C$
progressively. It starts with an empty canonical predicate (line 1) and inserts
one filter in each iteration (lines 2--13). Before insertion, it checks whether
$P^C$ is a canonical predicate (line 3) and terminates the loop if $P^C$ is
already valid. Otherwise, it picks the remaining filters that were not chosen in
earlier iterations as candidates (line 5) and inserts the filter that minimizes
the difference $\Delta(D-D_{P^C}-D_{p_i})$ at the highest magnitude into $P^C$
(lines 6--12). When homogeneity holds and $\Delta_i\leq \Delta(D-D_{P^C})$,
\A~\ref{alg:xplainer-avg} prunes $p_i$ according to
\Prop~\ref{prop:avg-inequality} (lines 7--8). Note that $\Delta_i$ is invariant
throughout the loop; thus it only needs to be queried once. In general cases
where homogeneity does not hold, \A~\ref{alg:xplainer-avg} has to enumerate all
possible filters in $\overline{P}$ (line 10). If we cannot obtain a valid
canonical predicate (i.e., a counterfactual cause to $\Delta$) after the loop,
\A~\ref{alg:xplainer-avg} terminates and outputs $\perp$, indicating that it
fails to find the optimal explanation within the attribute (line 15).
Empirically, we do not observe such rare cases. When the canonical predicate
$P^C$ is obtained, $\forall k=1,\cdots,|P^C|-1$, the top-k filters of
$P_k\subseteq P^C$ is a valid actual cause and the complement set
$\Gamma_k=P^C-P_k$ is a valid contingency. According to the termination
condition in the above loop (line 3), $\Delta(D-D_{P_k})> \epsilon$. In
addition, according to the definition of canonical predicate
$\Delta(D-D_{P^C})=\Delta(D-D_{P_k}-D_{\Gamma_k})\leq \epsilon$, $\Gamma_k$ is a
valid contingency to $P_k$. Therefore, we compute the approximated
responsibility $\hat{\rho_{P_k}}$ by using its lower bound deduced by $\Gamma_k$
(line 19). After enumerating each $k$, \A~\ref{alg:xplainer-avg} returns $P_k$
such that $\hat{\rho_{P_k}}-\sigma |P_k|$ is maximized (line 21). In summary,
the first loop (lines 2--14) is of quadratic complexity regardless of
homogeneity and the second loop is linear (lines 16--20). The total complexity
is $\mathcal{O}(m^2)$.

% \subsection{\todo{Implementation}}

% \begin{figure}[!htbp]
% \centering
% \includegraphics[width=0.92\columnwidth]{fig/workflow.pdf}
% \caption{Pipeline of \tool.}
% \label{fig:pipeline}
% \end{figure}

% \tool\ addresses the challenges of real-time effective explainable data analysis
% by decomposing the explanation search into a pipeline where most high-cost tasks
% are precomputed ahead of time. Only a few crucial tasks are conducted on the fly
% and are highly optimized. Holistically, \tool\ conducts 1) functional dependency
% detection, 2) skeleton discovery, and 3) causal orientation during
% initialization. Once it receives the real-time \diff\ from users, \tool\ 4)
% translates \diff\ into a list of candidates, 5) progressively searches through
% the candidates, and 6) ranks the discovered results based on multiple criteria.
% Finally, \tool\ presents the best-so-far results with various analytical
% semantics, such as explanations, coherence, and relevance, for explainable data
% analysis. 

\section{Evaluation}
\label{sec:eval}

In this section, we evaluate \tool\ to answer the following three research
questions (RQs): 

\begin{enumerate}[leftmargin=5mm]
  \item \textbf{RQ1: End-To-End Performance.}~How can \tool\ facilitate end
  users in explainable data analysis?
  \item \textbf{RQ2: \xlearn\ Evaluation.}~Does \xlearn\ effectively
  recover causal relations from observational data?
  \item \textbf{RQ3: \xplainer\ Evaluation.}~Does \xplainer\ accurately and
  efficiently yield explanations?\footnote{The correctness of
  \xtrans\ has been rigorously discussed in \S~\ref{subsec:csm}.}
\end{enumerate}

\subsection{Datasets \& User Study Setup}
\label{subsec:dataset}

To the best of our knowledge, there is no real-world benchmark with manually
labeled query/explanation pairs. To deliver a scientific evaluation, we conduct
experiments on \ding{172} public datasets collected from previous works,
\ding{173} real-world data collected from a production environment for user
study and human evaluation, and \ding{174} synthetic data with ground-truth
explanations. The detailed steps for generating synthetic datasets are given in
the \sm. We make necessary preprocessing before feeding to \tool\ (e.g., remove
missing values).

\parh{\ding{172} Flight Delay (FLIGHT).}~We use the flight delay dataset
from~\cite{salimi2017zaliql} to explore the causes of flight delays in US
airports. After preprocessing, the resulting dataset contains 17 variables,
including the weather conditions of departure airports (temperature, humidity,
visibility, rain, etc.), flight carrier, flight time (month, quarter, year, day
of week and hour) and two variables indicating flight delays, DelayMinute
(continuous) and Delay>15min (binary). 
% In particular, note
% that DelayMinute and Delay>15min form a functional dependency.

\parh{\ding{172} Hotel Booking (HOTEL).}~The hotel booking
dataset~\cite{antonio2019hotel} is frequently used for demonstrating data
analysis methods. It contains 119,390 observations from two hotels. Each
observation depicts the booking status (e.g., ``room type'', ``reservation
status'', and ``is canceled'') of a guest.

\parh{\ding{173} Web Service Behavior Dataset (WEB).}~The dataset is collected
from a web service's production environment. It contains 29 columns and 764
rows, where each row is a list of binary values. 
The first 28 columns describe user behaviors on the web service (e.g., whether
he clicks a specific button), which are collected by a logging module. The last
column indicates whether the user was blocked for publishing malicious content
(i.e., ``IsBlocked''), which is annotated by cybersecurity experts. These
behaviors are known to exhibit strong and clear causal relations, making it
appropriate for testing \tool\ in real-world scenarios.

\parh{\ding{174} Synthetic Data A (SYN-A).}~Ground-truth causal graphs are
unattainable in practice and it is common to generate random graphs and then
sample observational data from this graphical model. We generate MAGs with 10 to
150 variables (141 distinct scales in total). For each scale, we synthesize five
random graphs and the associated datasets, resulting in 705 ($141\times 5$)
datasets. \revision{Each dataset is injected with different amount of FDs.}

\parh{\ding{174} Synthetic Data B (SYN-B).}~We follow the approach in
Scorpion~\cite{wu2013scorpion} to synthesize datasets for assessing \xplainer.
Each dataset includes a valid \diff\ and a ground-truth explanation to this
difference. We generate 18 datasets with different difficulties.  

\parh{User Study Setup.}~\revision{In addition to the experiments that will be
launched shortly, we intend to determine the extent to which the results on
\textbf{WEB} is correct and reasonable. Nonetheless, rendering professional
judgments on explanations and causal claims require sufficient expertise in this
domain, which makes gathering a large number of participants difficult. In
this study, we recruit six domain experts for the \textbf{WEB} dataset; we
confirm each expert can evaluate the explanations and causal claims with
professionalism and high confidence.} We organize the user study as follows:

\begin{enumerate}[leftmargin=3mm]
  \item \textbf{Participant Education.}~We organize an education session for
  participants and demonstrate how to discern between causation and correlation.
  Then, a pilot study is conducted to confirm that participants have an adequate
  sense of causality.
  \item \textbf{Explanation Assessment.}~We raise four \diff\ and ask \tool\ to
  generate two explanations for each \diff. We then ask participants to give
  each explanation a score (between 0 to 5) based on their domain knowledge.
  \item \textbf{Causal Claim Assessment.}~Following~\cite{law2021causal}, we
  collect eight edges connected to ``IsBlocked'', transform these causal
  relations into human-comprehensible causal claims and ask participants to
  independently evaluate them (by labeling them as ``reasonable,'' ``not
  reasonable,'' or ``unsure'').
  \item \textbf{Follow-up Discussion.}~Participants explain their decision and
  discuss the aggregated results.
\end{enumerate}

\subsection{RQ1: End-To-End Performance}
\label{subsec:rq1}
We show that \tool\ generates plausible and intuitive explanations for diverse
datasets (\textbf{FLIGHT}, \textbf{HOTEL} and \textbf{WEB}) and invite experts
to assess the quality of explanations generated for \textbf{WEB}. In this
experiment, we manually discover noticeable differences to form \diff, and ask
\tool\ to supply the explanations. We also compare \tool's outputs with naive
correlation-based explanations. To ease presentation, we describe a \diff\ in
human-readable natural language in the following paragraphs.

\parh{FLIGHT \& HOTEL.}~We ask the following \diff\ on \ding{172}:
\begin{enumerate}
  \item FLIGHT: \textit{why \texttt{AVG(DelayMinute)} in May (24.95 min) is
  notably higher than the one in November (21.28 min)?}
  \item HOTEL: \textit{why \texttt{AVG(IsCanceled)} (cancellation rate) in July
  (0.37) is notably higher than the one in January (0.30)?}
\end{enumerate}

\begin{figure}[!t]
  \vspace{-5pt}
  \centering
\includegraphics[width=0.5\columnwidth]{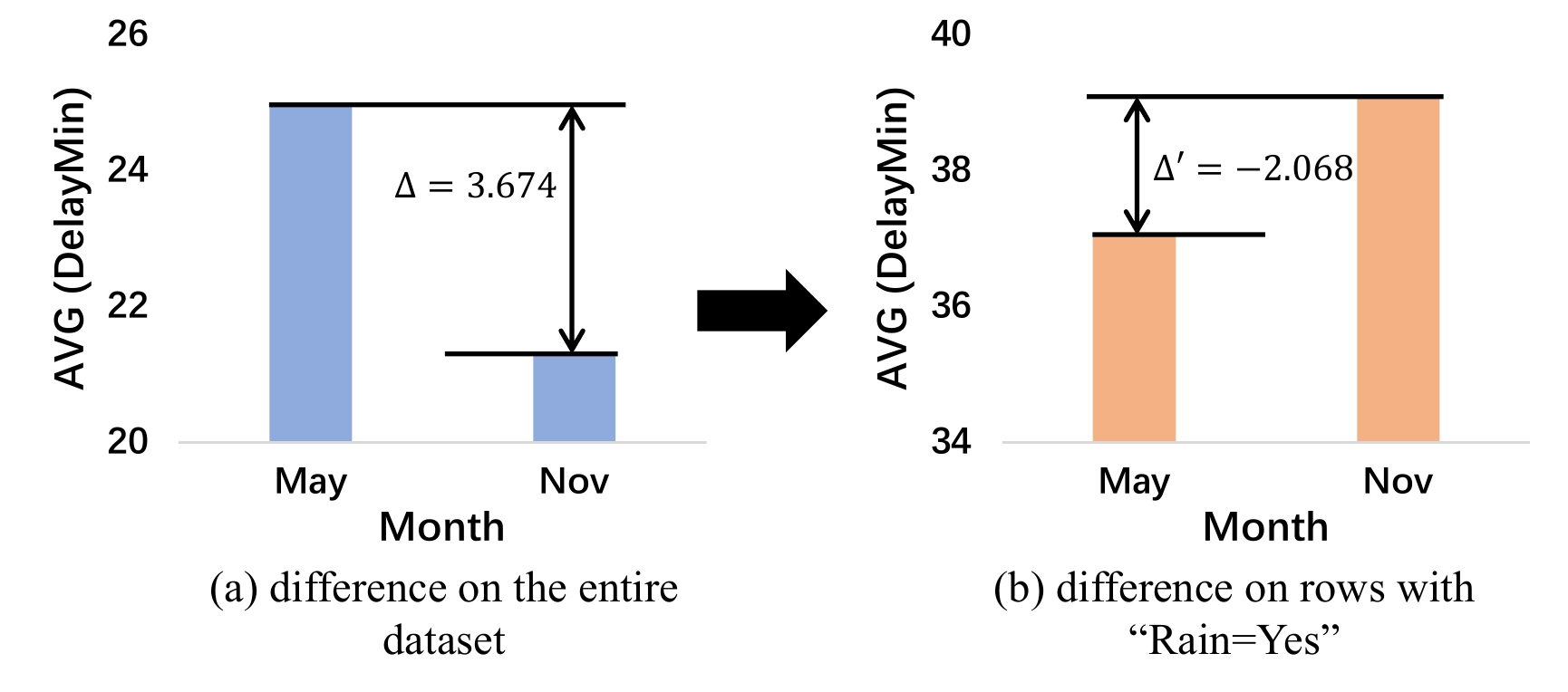}
\vspace{-10pt}
\caption{Explanation of \diff\ on the FLIGHT dataset.}
\label{fig:flight-diff-example}
\vspace{-5pt}
\end{figure}

For the first \diff, we observe that the duration of flight delay differs by
month, particularly for May and November, which motivates us to ask \tool\ for
explanations --- what the cause of the flight delay difference is. \tool\ first
learns a causal graph from data and identifies ``rain'' as a direct cause of
DelayMinute. Then, \xplainer\ finds that the difference is reversed
($\Delta=3.674$ vs. $\Delta'=-2.068$) when the condition ``rain=Yes'' is
enforced (\F~\ref{fig:flight-diff-example}). Thus, it returns ``rain=Yes'' as an
explanation. We interpret the explanation as correct because 1) rain is a
typical reason for flight delay, and 2) for most states, monthly precipitation
in May is usually higher than in November. Hence, when only counting the rainy
cases (by enforcing ``rain=Yes''), the difference is eliminated. 

For the second \diff, we observe that the cancellation rate varies by month of
arrival. For instance, the cancellation rate in July is 0.37, which is higher
than in January. Thus, we ask \tool\ for explanations. \tool\ identifies
``LeadTime'' (number of days between booking data and arrival date) as an
indirect cause of ``IsCanceled''. It also discovers that when enforcing
``LeadTime$\leq 133$'', the difference is reduced. This is intuitive. A longer
``LeadTime'' results in greater uncertainty about guests' future schedules,
leading to a higher cancellation rate. In January, LeadTime of most reservations
is less than 133 days (91\%). In contrast, there are far more early reservations
($>133$ days) in July (48\%), resulting in a higher cancellation rate. When
these early reservations are excluded, the difference becomes much smaller.

\begin{table}[!htbp]
  \vspace{-5pt}
  \centering
  \caption{\revision{Results of explanation assessment. E$i$ and P$i$ stand for
  the $i$th explanation and the $i$th participant, respectively.}}
  \vspace{-10pt}
  \resizebox{0.45\linewidth}{!}{
  \begin{tabular}{l|c|c|c|c|c|c|c|c}
    \hline
    & E1 & E2 & E3 & E4 & E5 & E6 & E7 & E8\\
    \thickhline
    P1 & 4 & 4 & 5 & 4 & 4 & 4&5 & 3\\\hline
    P2 & 4 & 4 & 4 & 4 & 3 & 4&3 & 4\\\hline
    P3 & 5 & 3 & 4 & 5 &3 &5 & 5& 5\\\hline
    P4 & 3 & 4 & 5 & 4 & 4& 3& 3& 4\\\hline
    P5 & 4 & 2 & 5 & 3 & 5& 4& 3& 3\\\hline
    P6 & 5 & 4 & 5 & 5 & 5& 4& 5& 5\\\thickhline
    mean & 4.16 & 3.50 & 4.67 & 4.17 & 4.00 & 4.00& 4.00& 4.00\\\hline
    std & 0.69 & 0.76 & 0.47 & 0.69 & 0.82 & 0.58& 1.00 & 0.82\\\hline

  \end{tabular}}
  \label{tab:web-human-assessment}
  \vspace{-5pt}
\end{table}

\parh{WEB.}~We report the results of the second phase in the user study (i.e.,
Explanation Assessment in \S~\ref{subsec:dataset}) in
\T~\ref{tab:web-human-assessment}. \revision{We view the results as encouraging,
since nearly all responses are positive ($\geq 3$). Moreover, the average scores
for seven out of eight explanations are $\geq 4$. We also investigated the
explanation with the lowest score (E2 in \T~\ref{tab:web-human-assessment}). {We
find this is counter-intuitive but reasonable in retrospect.} In the follow-up
session, the discussion among participants also confirmed our finding.} During
the assessments, experts find many explanations inspiring and insightful,
despite their familiarity with the dataset. It continuously increases their
knowledge and help them design a better criteria for detecting malicious
behavior.

\parh{Answer to RQ1:}~\textit{\tool\ shows a promising end-to-end performance in
explaining data differences. The user study validates that \tool\ achieves a
respectable level of agreement with experts.}

\subsection{RQ2: \xlearn\ Evaluation}
\label{subsec:xlearner-eval}

As a cornerstone of \tool, \xlearn\ is crucial to the effectiveness of the
entire pipeline. To answer \textbf{RQ2}, we run \xlearn\ on \textbf{SYN-A} which
has ground truth causal graphs and on a real-world dataset \textbf{WEB}. Since
\textbf{WEB} does not associate a ground-truth causal graph, we assess the
quality of causal relations by the user study.

\begin{table}[!htbp]
  % \vspace{-10pt}
  \centering
  \caption{Overall comparison between \xlearn\ and FCI.}
  % \vspace{-10pt}
  \resizebox{0.4\linewidth}{!}{
  \begin{tabular}{l|c|c|c}
    \hline
    \textbf{Algo.}& \textbf{F1-Score} & \textbf{Precision} & \textbf{Recall} \\
    \thickhline
    \xlearn & $0.88\pm 0.04$&$0.95\pm 0.03$ &$0.82\pm 0.06$\\\hline
    FCI & $0.72\pm 0.05$&$0.92\pm 0.04$ &$0.59\pm 0.06$\\
    \hline
  \end{tabular}}
  \label{tab:synthetic-comparison}
  % \vspace{-10pt}
\end{table}

\T~\ref{tab:synthetic-comparison} provides an overall comparison between
\xlearn\ and FCI on \textbf{SYN-A} datasets. We find that \xlearn\ is more
accurate than FCI in the presence of FDs. In particular, while FCI has
comparable precision, \xlearn\ has a much higher recall. This confirms our
discussion on the implications of FDs in \S~\ref{subsec:cdd}. The faithfulness
violations mislead FCI to incorrectly refute true edges (thus yield a lower
recall) while \xlearn\ is aware of such faithfulness violations and handles them
with the procedure in \S~\ref{subsec:cdd}.

\begin{figure}[!htbp]
  \vspace{-10pt}
  \centering
\includegraphics[width=0.65\columnwidth]{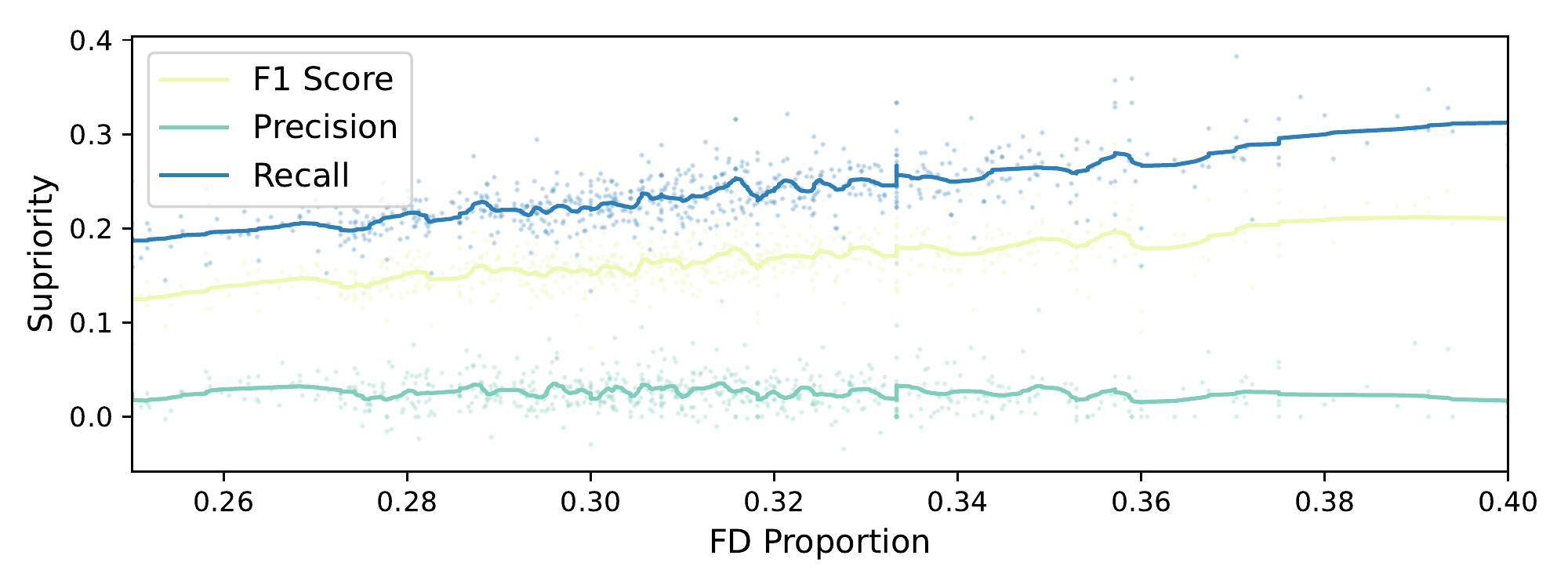}
\vspace{-10pt}
\caption{Comparison by FD Proportion. \revision{The x-axis is the proportion of
  FDs in the causal graph. The y-axis is the superiority (determined by
  subtracting the FCI's score from the \xlearn's score) of \xlearn\ over FCI.}}
\label{fig:synthetic-result}
\vspace{-10pt}
\end{figure}

\revision{Since \xlearn\ focuses primarily on FDs as the opposite of FCI, we
further study how varying proportions of FDs in the causal graph affect
\xlearn's performance. We report the superiority in terms of varied amounts of
FDs in \F~\ref{fig:synthetic-result}.} Overall, we observe an increasing trend
in \xlearn\ performance (particularly for F1 and recall) as the FD proportion
increases. \revision{More importantly, we observe that ``superiority'' increases
as FDs increase. Recall, as noted in the caption of
\F~\ref{fig:synthetic-result}, that the superiority is computed by subtracting
the FCI's score from the \xlearn's score. Thus, we interpret that \xlearn\
gradually outperforms the FCI algorithm with greater degree as the proportion of
FDs grows.}

In addition to the experiments on synthetic datasets, we also evaluate \xlearn\
with the \textbf{WEB} dataset (\S~\ref{subsec:dataset}). As aforementioned, this
real-world dataset lacks a ground-truth causal graph. Evaluating the accuracy of
an estimated causal graph is thus challenging, if not impossible. At this step,
we involve human experts to assess the correctness of the identified causal
relations in the third phase of our user study (i.e., Causal Claim Assessment in
\S~\ref{subsec:dataset}).

\begin{table}[!t]
  % \vspace{-10pt}
  \centering
  \caption{\revision{User study. C$i$ stands for the $i$th causal claim.}}
  % \vspace{-10pt}
  \resizebox{0.5\linewidth}{!}{
  \begin{tabular}{l|c|c|c|c|c|c|c|c}
    \hline
    & C1 & C2 & C3 & C4 & C5 & C6 & C7 & C8\\
    \thickhline
    \# Reasonable     & 6 & 4 & 4 & 6 & 6 & 4 & 5 & 5\\\hline
    \# Not Sure       & 0 & 2 & 1 & 0 & 0 & 0 & 1 & 1\\\hline
    \# Not Reasonable & 0 & 0 & 1 & 0 & 0 & 2 & 0 & 0\\\hline

  \end{tabular}}
  \label{tab:expert-feedback}
  % \vspace{-5pt}
\end{table}

We report the results of our user study in \T~\ref{tab:expert-feedback}:
\revision{first, out of 48 responses ($6 \text{~participants}~\times~8
\text{~questions}$), only three ($6.3\%$) suggest that the causal claims are
``Not Reasonable,'' while $40$ responses ($83.3\%$) mark the causal claims as
``Reasonable.''} It indicates that the causal relations identified by \xlearn\
correspond with expert knowledge in the majority of instances. Second, we
investigate the claims that have been deemed ``Not Reasonable'' or ``Unsure.''
Encouragingly, we find that a notable proportion of causal claims are
counter-intuitive yet \textit{correct}. For instance, one causal claim states
that ``malicious intent would lead to more frequent configuration changes than
benign intent.'' In the causal claim assessment phase, one expert deemed it
``Not Reasonable'' and presumed that malicious users would keep a default
configuration. In the follow-up session where they shared the independent
assessments, this participant was persuaded and confirmed this causal relation
as ``Reasonable.''

\parh{Answer to RQ2:}~\textit{As reflected by the carefully-designed
quantitative experiments and the user study, \xlearn\ generates plausible causal
graphs that are consistent with expert knowledge.}

\subsection{RQ3: \xplainer\ Evaluation}
\label{subsec:rq3}

Recall the descriptions of \textbf{SYN-B} in which different parameters result
in datasets with varying degrees of difficulty. In this experiment, we explore
the accuracy of \xplainer\ on \textbf{SYN-B}. 

\parh{Baseline.}~\revision{We compare \xplainer\ with three baselines, namely,
Scorpion~\cite{wu2013scorpion}, RSExplain~\cite{roy2014formal} and
BOExplain~\cite{lockhart2021explaining}, which use predicates as explanations.
Scorpion is an explanation engine for explaining outliers. It uses a metric
called influence score to rank explanations, which considers the effect of the
explanation between the outlier region and the hold-out region. RSExplain uses
the concept of intervention to measure the effectiveness of an explanation.
BOExplain is originally designed for explaining black-box machine learning
models. When explaining \diff, it employs the inference score and the Bayesian
optimization to find the optimal explanation. To launch an apple-to-apple
comparison, all baselines are enforced to search over a set of pre-defined
causal filters $\{p_1,\cdots,p_m\}$ derived from the generation procedure of
\textbf{SYN-B} (see details in \sm); all these filters have been confirmed to
constitute legitimate causal explanations.} 

\parh{Metric.}~We use the top-ranked explanation of each baseline as its optimal
explanation. We mark a method as ``N/A'' denoting timeouts (more than one hour
to process). We report the F1 Score of filters in the explanation over the
ground-truth explanation. 

\begin{table}[!htbp]
  \vspace{-5pt}
  \centering
  \caption{{\xplainer\ and baselines under various settings. \cmark\
  denotes that F1=1.0 and the best metric is \textbf{highlighted}.}}
  \vspace{-10pt}

  \resizebox{0.6\linewidth}{!}{
  \begin{tabular}{c|c|c|c|c|c|c|c}
    \thickhline
    \multicolumn{2}{c|}{\#Rows (Cardinality=10)} & 10K & 20K & 50K & 100K & 500K & 1M\\
    \hline

    \xplainer & F1 Score & \cmark & \cmark & \cmark & \cmark & \cmark & \cmark \\\cline{2-8}
    (\texttt{SUM})   & Time (sec.)& {\bf 0.004} & {\bf 0.005} & {\bf 0.007} & {\bf 0.010} & {\bf 0.017} & {\bf 0.019} \\\hline
    Scorpion & F1 Score & 0.5 & 0.5 & 0.5 & 0.5 & 0.5  & 0.8 \\\cline{2-8}
    (\texttt{SUM})   & Time (sec.)& 0.68 & 0.82 & 1.25 & 1.93 & 2.45 & 2.93 \\\hline
    RSExplain & F1 Score & 0.75 & 0.75 & 0.75 & 0.75 & 0.75 & 0.75 \\\cline{2-8}
    (\texttt{SUM})   & Time (sec.)& 0.68 & 0.83 & 1.25 & 1.94 & 2.44 & 2.90 \\\hline
    BOExplain & F1 Score & 0.8 & 0.8 & 0.5 & 0.5 & 0.5 & 0.8 \\\cline{2-8}
    (\texttt{SUM})   & Time (sec.)& 5.24 & 5.32 & 5.62 & 6.38 & 9.80 & 13.53 \\\hline \hline

    \xplainer & F1 Score & \cmark & \cmark & \cmark & \cmark & \cmark & \cmark \\\cline{2-8}
    (\texttt{AVG})   & Time (sec.)& {\bf 0.016} & {\bf 0.019} & {\bf 0.026} &{\bf  0.038} & {\bf 0.052} & {\bf 0.063 }\\\hline
    Scorpion & F1 Score & \cmark & \cmark & \cmark & \cmark & \cmark & \cmark \\\cline{2-8}
    (\texttt{AVG})   & Time (sec.)& 0.59 & 0.67 & 0.90 & 1.29 & 1.69 & 2.01 \\\hline
    RSExplain & F1 Score & 0.75 & 0.75 & 0.75 & 0.75 & 0.75 & 0.75 \\\cline{2-8}
    (\texttt{AVG})   & Time (sec.)& 0.58 & 0.66 & 0.90 & 1.28 & 1.68 & 1.95 \\\hline
    BOExplain & F1 Score & 0.86 & \cmark & 0.86 & \cmark & \cmark & 0.8 \\\cline{2-8}
    (\texttt{AVG})   & Time (sec.)& 5.33 & 5.37 & 5.56 & 6.56 & 8.67 & 12.62\\\hline \hline

    \multicolumn{2}{c|}{Cardinality (\#Rows=100k)} & 10 & 15 & 20 & 30 & 50 & 100\\
    \hline

    \xplainer & F1 Score & \cmark & \cmark & \cmark & \cmark & \cmark & {\bf 0.8 }\\\cline{2-8}
    (\texttt{SUM})   & Time (sec.)& {\bf 0.010} & {\bf 0.014} & {\bf 0.018} & {\bf 0.025} &{\bf  0.040} & {\bf 0.077} \\\hline
    Scorpion & F1 Score & 0.5 & 0.5 & 0.5 & N/A & N/A & N/A \\\cline{2-8}
    (\texttt{SUM})   & Time (sec.)& 1.96 & 16.50 & 75.72 & N/A & N/A & N/A \\\hline
    RSExplain & F1 Score & 0.75 & 0.75 & 0.75 & N/A & N/A & N/A \\\cline{2-8}
    (\texttt{SUM})   & Time (sec.)& 1.95 & 16.61 & 75.82 & N/A & N/A & N/A \\\hline
    BOExplain & F1 Score & \cmark & 0.86 & 0.86 & 0.46 & 0.27 & 0.15 \\\cline{2-8}
    (\texttt{SUM})   & Time (sec.)& 6.28 & 8.71 & 11.17 & 15.44 & 25.44  & 48.73 \\\hline \hline

    \xplainer & F1 Score & \cmark & \cmark & \cmark & \cmark & \cmark & \cmark \\\cline{2-8}
    (\texttt{AVG})   & Time (sec.)& {\bf 0.038} & {\bf 0.060} & {\bf 0.082} & {\bf 0.124} & {\bf 0.211} & {\bf 0.426} \\\hline
    Scorpion & F1 Score & \cmark & \cmark & \cmark & N/A & N/A & N/A \\\cline{2-8}
    (\texttt{AVG})   & Time (sec.)& 1.27 & 10.58 & 47.90 & N/A & N/A & N/A \\\hline
    RSExplain & F1 Score & 0.75 & 0.75 & 0.75 & N/A & N/A & N/A \\\cline{2-8}
    (\texttt{AVG})   & Time (sec.)& 1.28 & 10.59 & 47.91 & N/A & N/A & N/A\\\hline
    BOExplain & F1 Score & \cmark & 0.86 & 0.5 & 0.5 & 0.27 & 0.14 \\\cline{2-8}
    (\texttt{AVG})   & Time (sec.)& 5.87 & 8.23 & 10.44 & 15.00 & 24.35 & 46.35 \\    

    \thickhline
  \end{tabular}}
  \label{tab:rq3-size}
  \vspace{-5pt}
\end{table}

\parh{Different Dataset Sizes.}~\revision{To study the scalability of \xplainer,
we generate datasets of varying sizes and report the results in
\T~\ref{tab:rq3-size}. Overall, we observe that \xplainer\ is more accurate and
efficient than all baselines across all the studied settings. This is
encouraging and also reasonable, as \xplainer\ uses many distinct
characteristics of aggregation functions to optimize the search process, while
other methods primarily treat them as a ``black-box.'' Scorpion and BOExplain
often produce incomplete explanations, while RSExplain may frequently find extra
spurious filters. We presume that this is because the objective function of
Scorpion (and also BOExplain) is for explaining anomalies instead of \diff,
whereas RSExplain is primarily designed for data provenance. In contrast,
explanations provided by \xplainer\ are seen as consistent with the ground
truth.}

\revision{\xplainer\ is highly efficient particularly for high
cardinality regimes, while both Scorpion and RSExplain run out of time when the
cardinality exceeds 30 (see the bottom half of \T~\ref{tab:rq3-size}). BOExplain
uses Bayesian optimization to search for explanations, and its accuracy downgrades
as cardinality increases. Similarly, when iterating different \#Rows (the top
half of \T~\ref{tab:rq3-size}), \xplainer\ also exhibits highly encouraging
efficiency: \xplainer\ takes on average 0.06 seconds to explain \diff\ whereas
BOExplain (the second best) takes 13.17 seconds.
In sum, we interpret from \T~\ref{tab:rq3-size} that \xplainer\ delivers highly
encouraging accuracy and efficiency across different settings in comparison with
the baseline methods.}

\parh{Different $\mu^*-\mu$.}~\revision{The difference between $\mu^*,\mu$
indicates the magnitude of $\Delta$. To study the sensitivity of \xplainer, we
study how well \xplainer\ performs under varying differences and compare it with
baselines in \T~\ref{tab:rq3-mu}. To clarify, $\mu^*-\mu=5$ and $\mu^*-\mu=10$
denote two relatively more challenging settings in \T~\ref{tab:rq3-mu}, given
the very subtle differences.
On \texttt{SUM} aggregates, we find that all methods have difficulties in
identifying explanations in those two challenging settings; still, \xplainer\
yields the best results for both settings. Even on the most challenging setting
($\mu^*-\mu=5$), \xplainer\ finds highly accurate explanations whereas RSExplain
is less accurate.}

\begin{table}[!t]
  \vspace{-5pt}
  \centering
  \caption{{\xplainer\ and baselines with different $\mu^*-\mu$. \cmark\ denotes
  that the result is identical to the ground truth (F1=1.0).}}
  \vspace{-10pt}
  \resizebox{0.55\linewidth}{!}{
  \begin{tabular}{c|c|c|c|c|c|c}
    \hline
    $\mu-\mu^*$ & 5 & 10 & 15 & 30 & 50 & 100\\
    \thickhline
    \xplainer\ (\texttt{SUM}) & {\bf 0.86} & \cmark & \cmark & \cmark & \cmark & \cmark \\\hline
    Scorpion (\texttt{SUM}) & 0.50 & 0.50 & 0.50 & 0.50 & 0.50 & 0.50 \\\hline
    RSExplain (\texttt{SUM}) &  0.75 & 0.75 & 0.75 & 0.75 & 0.75 & 0.75\\\hline
    BOExplain (\texttt{SUM}) &  0.50 & 0.86 & 0.80 & 0.80 & 0.80 & \cmark\\\hline\hline
    \xplainer\ (\texttt{AVG}) & \cmark & \cmark & \cmark & \cmark & \cmark & \cmark \\\hline
    Scorpion (\texttt{AVG}) & 0.80 & \cmark & \cmark & \cmark & \cmark & \cmark \\\hline
    RSExplain (\texttt{AVG}) & 0.75 & 0.75 & 0.75 & 0.75 & 0.75 & 0.75 \\\hline
    BOExplain (\texttt{AVG}) &  0.80 & \cmark & 0.86 & 0.86 & 0.80 & \cmark\\\hline
  \end{tabular}}
  \label{tab:rq3-mu}
  % \vspace{-5pt}
\end{table}

\revision{On \texttt{AVG} aggregates, \xplainer\ and Scorpion both perform well
on identifying the ground-truth explanations; \xplainer\ is slightly better
particularly for the most challenging setting when $\mu^*-\mu=5$. Nevertheless,
RSExplain and BOExplain are less accurate on \texttt{AVG}. Overall, we conclude
that \xplainer\ is more robust to subtle data differences while all other
methods have difficulties in such challenging settings. We omit reporting the
processing time here, since it has already been evidently explored in
\T~\ref{tab:rq3-size}.}

\parh{Tightness of Approximation.}~In \S~\ref{subsec:xplainer}, we show the
approximation of minimal contingency for computing responsibilities under
\texttt{SUM} and \texttt{AVG}. In the following, we compare the tightness of the
responsibility $\hat{\rho}$ computed by $\overline{P}=P^C-P$ to the true
responsibility $\rho$ computed by the minimal contingency $P_{\min}$ via
brute-force search. The approximation error is computed as
$E=\frac{|\hat{\rho}-\rho|}{\rho}$. Recall that we craft three filters that form
the counterfactual cause in each dataset. In \texttt{SUM}, we can craft six
($3\choose 2$) actual causes from the three filters. In \texttt{AVG}, since the
canonical predicate of \texttt{AVG} only supports the first $k$ filters as
actual causes and the rest as contingencies, we pick the top-1 and top-2 filters
as two actual causes and repeat the experiments on three datasets ($2\times 3$
in total). On the six actual causes of \texttt{SUM} aggregates, we find that the
brute-force algorithm is $253.3\times$ slower than our approximated solution.
More importantly, the approximation error is highly negligible with an average
of $0.007$. We also observe that the approximation error on \texttt{AVG} is
slightly higher ($0.066$) and that our heuristics-based solution is $27.3\times$
faster. This result is reasonable, as the heuristics-based solution does not
provide guarantees of accuracy and requires more queries than \texttt{SUM}.

\parh{Answer to RQ3:}~\textit{\xplainer\ shows high scalability to large
datasets and also accurately generates explanations in very difficult settings.
On a mild cost of precision, two approximation solutions of \xplainer\
substantially improve efficiency.}

\section{Discussion}
\label{sec:discussion}

% \parh{\fixme{Single Dimension Explanation.}}~\tool\ primarily focuses on explanation
% over a single dimension. As pointed out in \S~\ref{subsec:prel-data}, \tool\ can
% be smoothly extended to multi-dimension by a Cartesian product on multiple
% dimensions. However, such multiple dimensions may impose complicated (and even
% obscure) causal semantics. Hence, we only consider a single dimension in \tool.

\parh{FD in Noisy Data.}~\revision{\tool\ only considers \textit{deterministic
FDs}. As illustrated in \Ex~\ref{ex:violation}, taking deterministic FDs into
account eliminates faithfulness violations. However, when the data is noisy, the
FDs may be stochastic (e.g., probabilistic interpretation of
FDs~\cite{zhang2020statistical}), which is currently not considered in \tool. 
%According to \Ex~\ref{ex:violation}, faithfulness
%violations are caused by the determinism of FDs; however, 
We clarify that considering only deterministic FDs deems a common setup shared
by relevant works in this field~\cite{ding2020reliable,mabrouk2014efficient}. It
remains unclear how noisy FDs may impact faithfulness. We leave this for future
exploration.}

\parh{Acquiring Causal Knowledge.}~Inferring causal relations is difficult.
Typically, it needs a combination of domain
knowledge~\cite{andrews2020completeness}, randomized
experiments~\cite{triantafillou2015constraint} and causal
discovery~\cite{spirtes2000causation,zhang2008completeness,ding2020reliable,ml4s,dai2021ml4c}.
\tool\ performs causal discovery from observational data due to its simplicity.
Nevertheless, we envision users of \tool\ can combine additional sources for
acquiring more accurate causal knowledge. In this paper, we explain several key
obstacles of applying causal discovery to real-world data, including causal
insufficiency~\cite{zhang2008completeness} and FD-induced faithfulness
violations~\cite{ding2020reliable,mabrouk2014efficient}. \xlearn, for the first
time, simultaneously addresses all of them.

\parh{Other Forms of Explanations.}~Currently, \tool\ employs predicates as the
content of explanations, which is general enough for common data analysis
scenarios. However, in some cases, explanations may be formed by the number of
records in a database~\cite{tableau} or counterbalances~\cite{miao2019going}.
Furthermore, when explaining changes in a whole time
series~\cite{chen2021tsexplain}, \tool\ may be not applicable. We leave
integrating \tool\ into these scenarios for future work.

\section{Related Work}
\label{sec:related}

\parh{Data Explanation.}~Explaining an unexpected query outcome in database is a
crucial phase in the lifecycle of data analysis. In general, an explanation aims
to provide certain forms of patterns that lead to the unexpected query outcome.
Such patterns may be a set of
predicates~\cite{wu2013scorpion,abuzaid2021diff,bailis2017macrobase,roy2014formal},
tuples~\cite{meliou2014causality}, or counterbalances~\cite{miao2019going}.
Scorpion is the most relevant work for \tool, which also provides explanations
to aggregated queries~\cite{wu2013scorpion}. In particular, it employs an
influence score to quantify explanations and features a set of optimizations to
reduce the cost of explanation search. Recently, many tools have attempted to
enhance explanations with additional knowledge (e.g., join tables) about the
underlying data. However, such additional knowledge does not imply causation ---
top ranked explanations could be rated low by human participants due to a lack
of causal semantics~\cite{li2021putting}. These observations evidently show the
necessity of integrating causality into XDA.

\parh{Causality in Database.}~Most works related to causality analysis in the
database is on the basis of Halpern's seminal framework on actual
causality~\cite{halpern2005causes,halpern2016actual}. It provides an elegant and
natural way to reason about input-output relations. Its results not only
highlight the output's cause, but also provide a contingency describing how it
is triggered. The adaption of actual causality in the database (i.e., DB
causality) is widely used for data provenance~\cite{meliou2010complexity}, data
explanation~\cite{roy2014formal} and
debugging~\cite{meliou2011tracing,fariha2020causality,yoon2016dbsherlock,ji2022perfce}.
However, it has limitations when applied alone. On the one hand, as noted
in~\cite{glavic2021trends}, DB causality does not necessarily imply \textit{true
causation}. Indeed, it assumes that causal knowledge is already known, and
focuses solely on quantitative explanations. On the other hand, considerable
adaptions are required to make it applicable to XDA scenarios, as discussed in
\S~\ref{subsec:xplainer}. We also notice other methods for quantifying
explanations, such as sufficient score, necessity score, and average causal
effect~\cite{salimi2020causal,watson2021local}. Despite their usefulness, we
design \xplainer\ on top of actual causality because it is more understandable
and general. Furthermore, without prior causal knowledge, none of these methods
can imply true causation.

\parh{XDA vs.~XAI.}~We note that XAI (explainable artificial intelligence) is
parallel and complementary to XDA. Through the lens of data analysis, XAI aims
to explain a prediction or
model~\cite{pradhan2022explainable,galhotra2021explaining}, while XDA enhances
EDA for understanding data facts. \revision{In addition, we also observe a line
of
research~\cite{wu2020complaint,flokas2022complaint,lin2022measuring,li2022training}
that identifies a subset of (training) data that is responsible for a
prediction. While this line of research shares similar output format with \tool,
it is essentially for explaining how model predictions are influenced by
training/test data, a scenario that is orthogonal to our research.}

\section{Conclusion}

This paper advocates XDA, a concept that ships comprehensive and in-depth
explainability toward EDA. XDA offers either causal or non-causal explanations
for EDA outcomes, from both quantitative and qualitative perspectives. We have
also presented the design of \tool, a production framework for XDA over
databases. Experiments and human evaluations reveal that \tool\ manifests highly
encouraging explanation capabilities. \xplainer\ has been integrated into
Microsoft Power BI to explain increase/decrease in data.

\begin{acks}
The authors would like to thank the anonymous reviewers for their valuable
comments and suggestions. The author would also like to thank Siwen Zhu, Haidong
Zhang, Zhitao Hou, Ruming Wang, Ziyu Wang, Long Ding, Kai Zhang, Jon Kay, and
Dingkun Xie for helpful discussions and all our participants in the user study
for their valuable feedback. The authors at HKUST were supported in part by RGC
RMGS under the contract RMGS22EG02.
\end{acks}

\printbibliography

\clearpage

\onecolumn
\section{Supplementary Material}

\subsection{\revision{Fast Causal Inference (FCI)
Algorithm~\cite{spirtes2000causation,zhang2008completeness}}}

We first introduce concepts and notations in addition to the one described in
\S~\ref{subsec:causal-discovery}. Then, we outline the FCI algorithm in
\A~\ref{alg:fci-sl} and \A~\ref{alg:fci-orient}.

\begin{definition}[Unshielded Triple]
    In a graph, a triple $(X, Y, Z)$ is an unshielded triple if $X$ and $Z$ are
    non-adjacent, $X$ and $Z$ are adjacent, and $Y$ and $Z$ are adjacent.
\end{definition}

\begin{definition}[Possible-D-SEP]
    In a graph, $\textit{Possible-D-SEP}(X,Y)$ is the set of nodes $Z$ such that
    there is an undirected path $\mathcal{P}$ between $X$ and $Z$ and for each
    subpath $S\starlinestar W\starlinestar T$ of $\mathcal{P}$ one of the
    following conditions holds.\footnote{\revision{$*$ denotes a wildcard
    endpoint; either $-$ (tail), $\rightarrow$ (arrowhead) or $\circline$
    (circle) can be matched.}}
    \begin{enumerate}
        \item $W$ is a collider; or,
        \item $W$ is not marked as a non-collider and $S, W, T$ are a triangle.
        (A triangle is a set of three nodes all adjacent to one another).
    \end{enumerate}
\end{definition}

\begin{definition}[Ext-D-SEP]
    In a graph, $\textit{Ext-D-SEP}(X,Y)$ is the union of
    $\textit{Possible-D-SEP}(X,Y)$ and $\textit{Possible-D-SEP}(Y,X)$.
\end{definition}

\begin{definition}[Discriminating Path]
    In a MAG, a path between $X$ and $Y$, $\mathcal{P}=(X, \cdots , W , V , Y)$,
    is a discriminating path for $V$ if
    \begin{enumerate}
        \item $\mathcal{P}$ includes at least three edges;
        \item $V$ is a non-endpoint node on $\mathcal{P}$, and is adjacent to
        $Y$ on $\mathcal{P}$; and
        \item $X$ is not adjacent to $Y$, and every node between $X$ and $V$
        is a collider on $\mathcal{P}$ and is a parent of $Y$.
    \end{enumerate}
\end{definition}

\begin{definition}[Uncovered Path]
    In a PMG (partial mixed graph), a path $\mathcal{P}=(V_0, \cdots , V_n)$ is
    said to be uncovered if for every $1\leq i\leq n-1$, $V_{i-1}$ and $V_{i+1}$
    are not adjacent, i.e., if every consecutive triple on the path is
    unshielded.
\end{definition}

\begin{definition}[Potentially Directed Path]
    In a PMG (partial mixed graph), a path $\mathcal{P}=(V_0, \cdots , V_n)$ is
    said to be potentially directed (abbreviated as p.d.) from $V_0$ to $V_n$ if
    for every $0\leq i\leq n-1$, the edge between $V_i$ and $V_{i+1}$ is not
    into $V_i$ or out of $V_{i+1}$.
\end{definition}

\begin{definition}[Circle Path]
    In a PMG (partial mixed graph), a circle path $\mathcal{P}=(V_0, \cdots ,
    V_n)$ is a special case of p.d.~path, where each edge on $\mathcal{P}$ is
    $\circlinecirc$.
\end{definition}

\begin{algorithm}[!htbp]
\caption{FCI-SL}
\label{alg:fci-sl}
\KwIn{Data $D$}
\KwOut{Skeleton $G$}
initialize a complete undirected graph $Q$ with $n$ nodes\;
$n\leftarrow 0$\;

\Repeat{for each ordered pair of adjacent nodes $X,Y$,
$\textit{Neighbor}(X)\setminus \{Y\}$ has less than $n$ neighbors}{

    \Repeat{all ordered pairs of adjacent nodes $X$ and $Y$ such that
    $\textit{Neighbor}(X)\setminus \{Y\}$ has cardinality $\geq n$ and all
    subsets $\bm{S}$ in $\textit{Neighbor}(X)\setminus \{Y\}$ have been tested
    for d-separation}{
    select an ordered pair of adjacent nodes $X,Y$ such that
    $\textit{Neighbor}(X)\setminus \{Y\}$ has cardinality $\geq n$, and a subset
    $\bm{S}\subseteq \textit{Neighbor}(X)\setminus \{Y\}$ of cardinality $n$,
    and, if $X \Perp Y \mid S$, delete the edge between $X$ and $Y$ from $Q$,
    and record $\bm{S}$ in $\textit{Sepset}(X,Y)$ and $\textit{Sepset}(Y,X)$;
    }
    $n\leftarrow n+1$\; 
}

let $F'$ be the undirected graph from the above step and orient each edge as
$\circlinecirc$\;
\ForEach{unshielded triple $(X, Y, Z)\in F'$}{
    \If{$Y$ is not in $\textit{Sepset}(X,Z)$}{
        orient $X\starlinestar Y\starlinestar Z$ as $X\stararrow Y \arrowstar Z$\;
    }
}
\ForEach{adjacent pair $(X, Y)\in F'$}{
    \If{there exists a subset $\bm{S}\subseteq \textit{Ext-D-SEP}(X,Y)$ such that $X\Perp Y\mid \bm{S}$}{
        delete the edge between $X$ and $Y$ from $F'$\;
    }
}
let $G$ be the undirected graph of $F'$\;
\Return{$G, \textit{Sepset}$}
\end{algorithm}

\begin{algorithm}[!htbp]
\caption{FCI-Orient}
\label{alg:fci-orient}
\KwIn{Skeleton $S$, Sepset \textit{Sepset}}
\KwOut{PAG $G$}

let $G$ be the graph sharing all adjacencies of $S$ and orient each edge as
$\circlinecirc$\;
\ForEach{unshielded triple $(\alpha , \beta, \gamma)\in G$}{
    \If{$\beta$ is not in $\textit{Sepset}(\alpha,\gamma)$}{
        orient $\alpha\starlinestar \beta\starlinestar \gamma$ as $\alpha\stararrow\beta \arrowstar \gamma$\;
    }
}
\Repeat{none of the orientation rules applies}{

$\mathcal{R}1$: if $\alpha \stararrow\beta \circstar \gamma $, and $\alpha$ and
$\gamma$ are not adjacent, then orient the triple as $\alpha \stararrow \beta
\rightarrow \gamma$\;

$\mathcal{R}2$: if $\alpha \rightarrow\beta \stararrow \gamma $ or $\alpha
\stararrow\beta \rightarrow \gamma$, and $\alpha \starcirc \gamma$, then orient
$\alpha \starcirc \gamma$ as $\alpha \stararrow \gamma$\;

$\mathcal{R}3$: if $\alpha \stararrow\beta \arrowstar \gamma$, $\alpha
\starcirc \theta \circstar \gamma$, and $\alpha$ and $\gamma$ are not adjacent,
and $\theta \starcirc \beta$, then orient $\theta \starcirc \beta$ as $\theta
\stararrow \beta$\;

$\mathcal{R}4$: if $u=(\theta, \cdots , \alpha , \beta, \gamma)$ is a
discriminating path between $\theta$ and $\gamma$ for $\beta$, and
$\beta\circstar \gamma$; then if $\beta\in\textit{Sepset}(\theta,\gamma)$,
orient $\beta\circstar \gamma$ as $\beta\rightarrow \gamma$; otherwise orient
the triple $(\alpha , \beta, \gamma)$ as $\alpha\leftrightarrow \beta
\leftrightarrow \gamma$\;
}

\Repeat{none of the orientation rules applies}{

$\mathcal{R}5$: for every (remaining) $\alpha\circlinecirc\beta$, if there is an
uncovered circle path $p=(\alpha,\gamma,\cdots,\theta,\beta)$ between $\alpha$
and $\beta$ s.t.~$\alpha$ and $\beta$ are not adjacent, and $\gamma$ and
$\theta$ are not adjacent, then orient $\alpha\circlinecirc\beta$ and every edge
on $p$ as undirected edges (—)\;

$\mathcal{R}6$: if $\alpha-\beta\circstar\gamma$ ($\alpha$ and $\gamma$ may or
may not be adjacent), then orient $\beta\circstar\gamma$ as
$\beta\linestar\gamma$\;

$\mathcal{R}7$: if $\alpha-\beta\circstar\gamma$ , and $\alpha$ and $\gamma$ are
not adjacent, then orient $\beta\circstar\gamma$ as $\beta\linestar\gamma$\;

$\mathcal{R}8$: if $\alpha\rightarrow\beta\rightarrow\gamma$ or
$\alpha\linecirc\beta\rightarrow\gamma$, and $\alpha \circarrow\gamma$, orient
$\alpha \circarrow\gamma$ as $\alpha \rightarrow\gamma$\; 

$\mathcal{R}9$: if $\alpha \circarrow\gamma$ and
$p=(\alpha,\beta,\theta,\cdots,\gamma)$ is an uncovered p.d. path from $\alpha$
to $\gamma$ such that $\beta$ and $\gamma$ are not adjacent, then orient $\alpha
\circarrow\gamma$ as $\alpha \rightarrow\gamma$\;

$\mathcal{R}10$: suppose $\alpha \circarrow\gamma$, $\beta
\rightarrow\gamma\leftarrow \theta$, $p_1$ is an uncovered p.d. path from
$\alpha$ to $\beta$ and $p_2$ is an uncovered p.d. path from $\alpha$ to
$\theta$. Let $\mu$ be the node adjacent to $\alpha$ on $p_1$ ($\mu$ could be
$\beta$), and $\omega$ be the node adjacent to $\alpha$ on $p_2$ ($\omega$
could be $\theta$). If $\mu$ and $\omega$ are distinct, and are not adjacent,
then orient $\alpha \circarrow\gamma$ as $\alpha \rightarrow\gamma$\;
}

\Return{$G$}

\end{algorithm}

\color{black}
\everymath{\color{black}}
\everydisplay{\color{black}}

\subsection{Principle of Explainability under \texttt{AVG} and \texttt{SUM}}

In \texttt{AVG}, we notice that $\texttt{AVG}_M(s_1)\approx \mathbb{E}(M\mid
F=f_1)$ and $\texttt{AVG}_M(s_2)\approx \mathbb{E}(M\mid F=f_2)$ asymptotically.
$\texttt{AVG}_M(s_1\land X=x)\approx \mathbb{E}(M\mid F=f_1,
X=x)=\mathbb{E}(M\mid F=f_1)=\texttt{AVG}_M(s_1)$. Similarly,
$\texttt{AVG}_M(s_2\land X=x)\approx\texttt{AVG}_M(s_1)$. Hence,
$\Delta(D)\approx \Delta(D_{X=x})$.

In \texttt{SUM}, we notice that $\texttt{SUM}=\texttt{COUNT}\times\texttt{AVG}$
and $\texttt{COUNT}_M(s_1)=N\times P(F=f_1), \texttt{COUNT}_M(s_2)=N\times
P(F=f_2)$, where $N$ is a constant indicating the number of rows in the data.
Then, $\Delta=N(P(F=f_1)\mathbb{E}(M\mid F=f_1)-P(F=f_2)\mathbb{E}(M\mid
F=f_2))$. $\Delta(D_{X=x})=N(P(F=f_1, X=x)\mathbb{E}(M\mid F=f_1,
X=x)-P(F=f_2,X=x)\mathbb{E}(M\mid F=f_2, X=x)) =N(P(F=f_1, X=x)\mathbb{E}(M\mid
F=f_1)-P(F=f_2, X=x)\mathbb{E}(M\mid F=f_2))$.

\subsection{Proof of \Thm~\ref{thm:single-sink}}

\begin{lemma}
    \label{lem:1}
    If $X\fdarrow Y$, then $Y\not\!\perp\!\!\!\perp X$, and for any other variable set $Z$, $Z\Perp Y\mid X$.
\end{lemma}

\begin{proof}
    Because $X\fdarrow Y$, $P(Y\mid X)=I_{Y=f(X)}$. Since $|X|, |Y| >1$, the following inequality holds.
    \begin{equation}
        P(X Y)=P(X) P(Y \mid X)=P(X) I_{Y=f(X)} \neq P(X) P(Y)
    \end{equation}
    Therefore, $Y\not\!\perp\!\!\!\perp X$. Given a variable set $Z$, 
    \begin{equation}
        \begin{aligned}
            P(Y Z \mid X)&=\frac{P(X Y Z)}{P(X)}\\
                        &=\frac{P(X Z) I_{Y=f(X)}}{P(X)}\\
                        &=P(Z \mid X) I_{Y=f(X)}\\
                        &=P(Z \mid X) P(Y \mid X)
        \end{aligned}
    \end{equation}
    Therefore, $Z\Perp Y\mid X$.
\end{proof}

\begin{lemma}
    \label{lem:2}
    If $X\fdarrow Y$ and $Z\Perp X\mid W$, then $Z\Perp Y\mid W$, where $Z$ and $W$ are two disjoint variable set 
    other than $X$ and $Y$.
\end{lemma}

\begin{proof}
    Let $\pi_y=\{x\mid f(x)=y\}$. Therefore, $\{X\in \pi_y\}=\{Y=y\}$. Since $Z\Perp X\mid W$,
    \begin{equation}
        \begin{aligned}
        P(Y=y,Z\mid W)&=P(X\in \pi_y, Z\mid W)\\
        &=P(X\in \pi_y\mid W)P(Z\mid W)\\
        &=P(Y=y\mid W)P(Z\mid W)
        \end{aligned}
    \end{equation}
    Therefore, $Z\Perp Y\mid W$.
\end{proof}

\begin{lemma}
    \label{lem:3}
    In a MAG $\mathcal{G}$ over variables $\{X_1,\cdots,X_n\}$, if
    $X_1\rightarrow X_2$ and no other edges connect $X_2$ with other vertices,
    when $X_2\sep Y\mid U$ for any $Y\in \{X_3,\cdots,X_n\}, U\subset
    \{X_3,\cdots,X_n\}, Y\notin U$, $X_1\sep Y\mid U$. 
\end{lemma}

\begin{proof}
    If $X_2\sep Y\mid U$, according to the definition of m-separation, each path between $X_2$ and $Y$ 
    satisfies one of the following two conditions: 1) there exists $X_p\rightarrow X_s\rightarrow X_q$,
    $X_p\leftarrow X_s\rightarrow X_q$, or $X_p\leftarrow X_s\leftarrow X_q$ where $X_s\in U$. 2) there 
    exists $X_p\rightarrow X_s\leftarrow X_q$ where $X_s$ or any of its descendants does not belong to $U$.
    Because $X_1\notin U$, $X_1\neq X_s$. For the first case, $X_s$ blocks $X_1$ and $Y$ given that there
    is only one edge connecting to $X_2$. For the second case, $X_1$ cannot be a collider, because 
    $X_1\rightarrow X_2$. Therefore, $U$ also blocks $X_1$ and $Y$ on this path. In summary, 
    $X_1\sep Y\mid U$.
\end{proof}

With above lemmas, now we prove \Thm~\ref{thm:single-sink}.

\begin{proof}
    We first construct a MAG $\mathcal{G}$ on the top of the skeleton, by adding
    an arrowhead from $X_1$ to $Z$ ($\mathcal{G}_2$). Because $\mathcal{S}_1$ is
    learnt from $\{X_1,\cdots,X_n\}$ where faithfulness assumption holds, this
    part is harmonious. For $Z$, there are two types of m-separation in
    $\mathcal{G}$. Here, we prove that each type of m-separation satisfies GMP
    to data distribution $P_V$.

    Type 1: $Z\sep X_{j\neq 1}\mid X_1$
    
    According to \Lem~\ref{lem:1}, $X_1\fdarrow Z$ implies that $Z\Perp X_{j\neq 1}\mid X_1$.

    Type 2: $Z\sep X_j\mid U (j\neq 1, U\cap \{X_1,X_j\}=\emptyset)$

    By \Lem~\ref{lem:3}, $Z\sep X_j\mid U$ implies that $X_1\sep X_j\mid U$. Because $S_1$
    is harmonious, $X_1\sep X_j\mid U$ implies that $X_1\Perp X_j\mid U$. According to 
    \Lem~\ref{lem:2}, $X_1\fdarrow Z$ and $X_1\Perp X_j\mid U$ imply $Z\Perp X_j\mid U$.

    Minimality is satisfied, since removing edge from $X_1$ to $Z$ will make $X_1\Perp Z$ which 
    contradicts $P_V$.
\end{proof}

\subsection{Proof of \Thm~\ref{thm:alg}}

\begin{proof}
We prove \Thm~\ref{thm:alg} by mathematical induction. In the sense, \A~\ref{alg:cdd} returns
a harmonious skeleton when $\fdgraph$ has arbitrary number of non-root vertices. Denote the number of
non-root vertices as $s\geq0$.

\textbf{Base Case.}
When $s=0$, the skeleton is harmonious because all variables are under faithfulness assumptions.
% When $s=1$, by \Thm~\ref{thm:single-sink}, the skeleton is harmonious.

\textbf{Induction.}
Suppose the returned skeleton is harmonious for $\fdgraph$ when $s=n$. Now we
prove the skeleton is still harmonious when we add a vertex $X'$ to a vertex set
$\bm{X}\subseteq \fdgraph.V$. Denote the new FD-induced graph as
$\fdgraph'$, the skeleton of $\fdgraph$ as $\mathcal{S}$, and the skeleton for
$\fdgraph'$ as $\mathcal{S}'$. According to \A~\ref{alg:cdd},
$\mathcal{S},\mathcal{S}'$ share the same vertices and edges (except $X'$).
Given that $\mathcal{S}$ is harmonious, $\mathcal{S}'$ can be decomposed into
two subgraphs, i.e., $\mathcal{S}_1,\mathcal{S}_2$, where
$\mathcal{S}_1=\mathcal{S}$ and $\mathcal{S}_2$ corresponds to $X'$ and one edge
from $X'$ to one of $\bm{X}$. Since $X''\fdarrow X'$, by
\Thm~\ref{thm:single-sink}, $\mathcal{S}'$ is harmonious.

By the principle of mathematical induction, \Thm~\ref{thm:alg} holds.

\end{proof}

\subsection{FCI Rules on FD-related Edges}

\begin{figure}[!htbp]
\centering
\includegraphics[width=0.4\columnwidth]{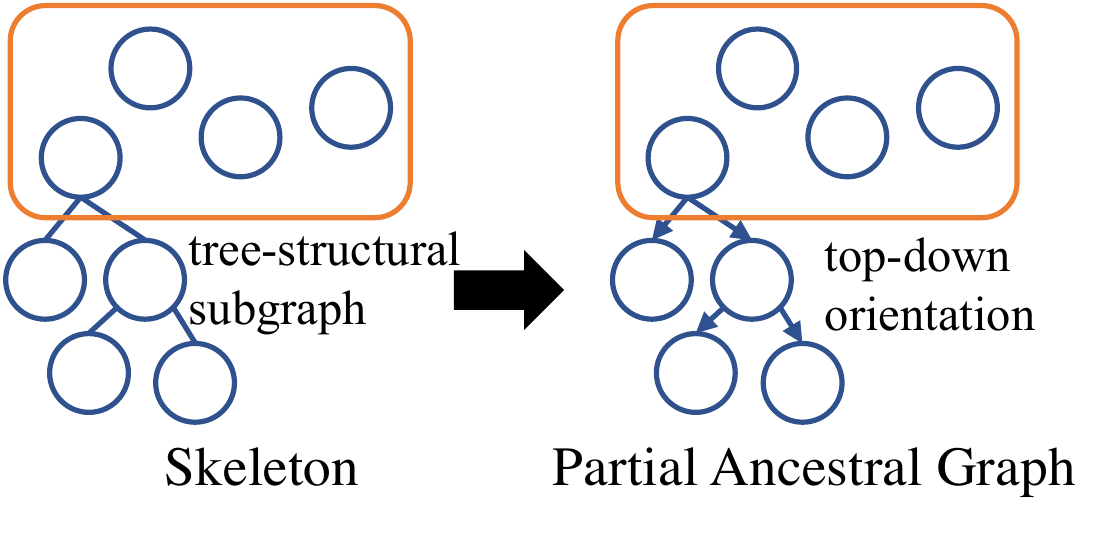}
\vspace{-10pt}
\caption{Orientation on tree-structural subgraph.}
\label{fig:orientation}
\end{figure}

Since we only consider one-to-one and one-to-many FDs, FD-induced vertices only
have at most one parent node in $\fdgraph$. Therefore, it forms
tree-structural subgraphs in the skeleton starting from the root vertices of
$\fdgraph$ (e.g., the left-hand side skeleton shown in
\F~\ref{fig:orientation}). Intuitively, it would be ideal if the orientation
rules allow us to orient the tree-structural subgraphs in a top-down manner
(e.g., the right-hand side of \F~\ref{fig:orientation}). In particular, if there is an
edge heading to the root node of the tree-structural subgraph, we can assign
$\rightarrow$ ($\rightarrow$ is defined in \T~\ref{tab:pag-edges}) as the edge
from the root node to its child vertices and propagate the directions to the
leaf node by applying Rule 1 of FCI recursively. 

\begin{definition}[Rule 1 of FCI~\cite{zhang2008completeness}]
    \label{def:rule1}
    If $X\stararrow Y\circarrow Z$ and $X,Z$ are non-adjacent, then orient 
    $Y\rightarrow Z$.~\footnote{``$*$'' is used as a wildcard
    symbol accepting either $\rightarrow,\leftrightarrow,\circarrow$.}
\end{definition}

However, it is not necessarily attainable if we cannot identify the direction on
the edge heading to the root node (i.e., $Y$ in \D~\ref{def:rule1}). As a
result, we cannot derive any directions on the subgraphs. That is, all edges on
the subgraph are $\circlinecirc$, which denote edges with unknown directions.
Such ambiguous cases are surely undesirable in XDA scenarios. 

\subsection{ANM on FD-related Edges}

We note that, from viewpoint of conditional independence, functional dependency
does not imply causal directions in all cases. However, we present that the
counterexamples of such cases (i.e., contrasting direction between causal
relation and functional dependency) are very rare in practice, if at all
possible. Recall the theorem of ``Identifiability of discrete ANMs''
in~\cite{peters2017elements}.

\begin{theorem}
    \label{thm:anm-identifiability}
    Assume that a distribution $P_{X,Y}$ allows for an ANM $Y = f(X) + N_Y$ from $X$ to $Y$ 
    and that either $X$ or $Y$ has finite support. $P_{X,Y}$ allows for an ANM from $Y$ to $X$ 
    if and only if there exists a disjoint decomposition $\bigcup_{i=0}^{l} C_{i}=\text{supp} X$, 
    such that the following conditions a), b), and c) are satisfied:
    
    a) The $C_i$’s are shifted versions of each other
    \begin{equation}
        \forall i\exists d_i >0: C_i=C_0+d_i
    \end{equation}
    and $f$ is piecewise constant: $f\mid C_i \equiv c_i\forall i$.
    
    b) The probability distributions on the $C_i$s are shifted and scaled versions of
    each other with the same shift constant as above: For $x\in C_i$, $P(X = x)$
    satisfies 
    \begin{equation}
        P(X=x)=P(X=x-d_i)\cdot \frac{P(x\in C_i)}{P(x\in C_0)}
    \end{equation}
    
    c) The sets $c_i +\text{supp} N_Y := \{c_i +h : P(N_Y = h) > 0\}$ are disjoint sets.
\end{theorem}

By condition c), since $N_Y=0$ by functional dependency, $\nexists x_i\in C_i,
x_j\in C_j$ s.t. $f(x_i)=f(x_j)$. In other word, $X$ are decomposed by the value
of $f(x)$ and each $c_i\in f(X)$ forms a disjoint subset $C_i$ of $X$. To admit
condition a), each $C_i$ is a shifted version of others. Therefore, each $C_i$
should at least have an equal cardinality. The above two conditions imply that
each $y\in Y$ corresponds to equal size of $x\in X$. Consider the aforementioned
example where $\text{Country}\fdarrow \text{Continent}$. To satisfy the
condition, each continent must have equal number of countries. Furthermore, by
condition b), the probability of a country $x$ in continent $c_i$ in the
database can be exactly scaled by $\frac{P(x\in C_i)}{P(x\in C_0)}$ the
probability of another shifted country $x-d_i$. The conditions together exhibits
a very rare scenario. Therefore, we ignore such cases in practice.

As consistent with~\cite{peters2017elements}, we consider it reasonable to infer
that direction of ANM as causal. Therefore, we hypothesis that functional
dependencies in $\fdgraph$ intrinsically imply causal directions.

\subsection{Proof of \Prop~\ref{prop:positive}}

\begin{proof}
    Suppose there exists an optimal explanation $P^*$ such that $\exists p_i\in
    P^*, \Delta_i \leq 0$. Let $P'=\{p\mid p\in P^*, \Delta_p>0\}$ and $\Gamma$
    be the minimal contingency of $P^*$. 
    
    For \texttt{SUM}, since $\Delta(D_{P'})\geq \Delta(D_{P^*})$,
    $\Delta(D)-\Delta(D_\Gamma)-\Delta(D_{P'}) <
    \Delta(D)-\Delta(D_\Gamma)-\Delta(D_{P^*}) \leq \epsilon$. 
    
    For homogeneous \texttt{AVG},
    \begin{equation}
        \begin{aligned}
            \Delta(D-\Gamma-P')&=\frac{\sum_{p_i\in \{p_1,\cdots,p_m\}-\Gamma-P'} a_i\Delta_i}{A_{D-\Gamma-P'}}\\
            &<\frac{\sum_{p_i\in \{p_1,\cdots,p_m\}-\Gamma-P^*} a_i\Delta_i}{A_{D-\Gamma-P^*}}\\
            &=\Delta(D-\Gamma-P^*)\leq \epsilon
        \end{aligned}
    \end{equation}
    
    Hence, $\Gamma$ is also a valid contingency for $P'$. $\rho_{P'}\geq
    \rho_{P^*}$ and $|P'|<|P^*|$ contradict the fact that $P^*$ is optimal. 
\end{proof}

\subsection{Proof of \Prop~\ref{prop:completeness}}

\begin{proof}
    When there are more than one optimal explanations, let $P'$ be the one with
    the smallest predicate size (i.e., $|P|$) and $\Gamma'$ be the corresponding
    minimal contingency. Otherwise, let $P'$ and $\Gamma'$ be the optimal
    explanation and corresponding minimal contingency, respectively. If
    $|P'|\geq |P^C|$, given that $\rho_{P'}\leq 1$ and $\rho_{P^C}=1$,
    $\rho_{P'}-\sigma\rho_{P'}\leq\rho_{P^C}-\sigma\rho_{P^C}$. $P^C$ is at
    least as optimal as $P'$. If $|P'|<|P^C|$, then let $P''$ be a predicate by
    replacing all non-canonical filters in $P'$ with canonical ones. Then,
    $1-d_{P''}\leq 1-d_{P'}\leq \epsilon'$. $\Gamma'$ is then a valid
    contingency for $P''$. Therefore, $\rho_{P''}\geq \rho_{P'}$ and
    $|P''|=|P'|$. $\rho_{P''}-\sigma\rho_{P''}\geq\rho_{P'}-\sigma\rho_{P'}$.
    $P''\subset P^C$ is at least as optimal as $P'$. In summary, there must
    exist an optimal explanation $P''\subseteq P^C$.
\end{proof}

\subsection{Proof of \Thm~\ref{thm:complement-set-contingency}}

\begin{proof}
    Let $m_k=\sum_{i=1}^k d_{p_i}$. According to the definition of canonical
    predicate, $1-m_j\leq \epsilon'<1-m_{j-1}$. This inequality implies that
    $1-(m_j-d_{p_j}) + d_{p_j}\leq\epsilon'<1-(m_j-d_{p_j})$. Since
    $\overline{P}\subset P^C$ and $d_{p_1}\leq \cdots\leq d_{p_j}$,
    $d_{\overline{P}}\geq d_{p_j}$. The following inequalities hold.
    \begin{equation}
        \begin{gathered}
            \epsilon'\geq 1-(m_{j}-d_{\overline{P}})-d_{\overline{P}} = 1-d_P-d_{\overline{P}}\\
            \epsilon'< 1-(m_{j}-d_{p_{j}})\leq 1-(m_{j}-d_{\overline{P}})=1-d_P
        \end{gathered}
    \end{equation}
    Since $1-d_P-d_{\overline{P}}\leq\epsilon'<1-d_P$,
    \Thm~\ref{thm:complement-set-contingency} holds.
\end{proof}

\subsection{Proof of \Thm~\ref{thm:approx}}

\begin{proof}
$\rho_P=\frac{1}{1+\min(|\Gamma|_W)}\geq
\frac{1}{1+|\overline{P}|_W}=\frac{1}{1+m_{j}-d_P}$. Furthermore, since
$1-d_P-d_\Gamma\leq\epsilon'$, $d_\Gamma\geq 1-d_P-\epsilon'$ and $\rho_P\leq
\frac{1}{2-d_P-\epsilon'}$. Thus, we derive the lower and upper bounds of
$\rho_P$. When assuming $d_P\ll m_j$ and $0<m_j\leq 1$,
$\frac{1}{1+m_{j}-d_P}=\frac{1+m_{j}+d_P}{(1+m_{j})^2-d_P^2}\approx\frac{1+m_{j}+d_P}{(1+m_{j})^2}$.
The approximation error rate
$E=\frac{\frac{1}{1+m_{j}-d_P}-\frac{1+m_{j}+d_P}{(1+m_{j})^2}}{\frac{1}{1+m_{j}-d_P}}
=\frac{d_P^2}{(1+m_{j})^2}< \frac{m_{j}^2}{(1+m_{j})^2}\leq 0.25$.
\end{proof}

\subsection{Proof of \Prop~\ref{prop:avg-inequality}}

To prove \Prop~\ref{prop:avg-inequality}, we first introduce the following
notations and propositions.

\noindent\textbf{Notations.}~Given \diff\ $\Delta$ and corresponding sibling
subspaces $s_1,s_2$, let $x_i,y_i$ be $agg_M(D_{s_1\cap p_i})$ and
$agg_M(D_{s_1\cap p_i})$, respectively. We also let $a_i,b_i$ be $|D_{s_1\cap
p_i}|$ and $|D_{s_1\cap p_i}|$ denoting the rows of $D_{s_1\cap p_i}$ and
$D_{s_2\cap p_i}$, respectively, and $A_D,B_D$ be $|D_{s_1}|$ and $|D_{s_2}|$.
Then, it is obvious that $\Delta(D_{p_i})=x_i-y_i$ and we use $\Delta_i$ as a
shorthand for $\Delta(D_{p_i})$. For \texttt{AVG} and $\Delta(D_P)=\sum_{p_i\in
P} \Delta_i$, $\Delta(D)$ can be represented in the form of $\sum_1^m (\frac{a_i
x_i}{A_D}-\frac{b_i y_i}{B_D})$. 

\begin{proposition}
    \label{prop:homogeneous}
    If a pair of sibling subspaces are homogeneous on $X$, then
    $\frac{a_1}{b_1}=\cdots=\frac{a_m}{b_m}$.
\end{proposition}

\begin{proof}
    For homogeneous \texttt{AVG}, $X,F$ are m-separated given $\boldsymbol{B}$.
    Hence, $X\Perp F\mid \boldsymbol{B}$ and $P(X,F\mid \boldsymbol{B})=P(X\mid
    \boldsymbol{B})P(F\mid \boldsymbol{B})$. Recall that $a_i=P(X=x_i,F=f_1\mid
    \boldsymbol{B}=\boldsymbol{b})$ and $b_i=P(X=x_i,F=f_2\mid
    \boldsymbol{B}=\boldsymbol{b})$. We have $\frac{a_i}{b_i}=\frac{P(F=f_1\mid
    \boldsymbol{B}=\boldsymbol{b})}{P(F=f_2\mid \boldsymbol{B}=\boldsymbol{b})}$ 
    which is a constant and invariant with respect to $i$.
\end{proof}

\begin{proposition}
    \label{prop:additive}
    If $P_1$ and $P_2$ are two disjoint predicates on the same attribute,
    for homogeneous \texttt{AVG}, $\Delta(D_{P_1}+D_{P_2})< \Delta(D_{P_1}) +
    \Delta(D_{P_2})$.
\end{proposition}

\begin{proof}
    For homogeneous \texttt{AVG}, $\Delta(D_{P_1})=\frac{\sum_{p_i\in P_1} a_i
    \Delta_i}{A_{P_1}}$, $\Delta(D_{P_2})=\frac{\sum_{p_i\in P_2} a_i
    \Delta_i}{A_{P_2}}$ and $\Delta(D_{P_1}+D_{P_2})=\frac{\sum_{p_i\in P_1\cup
    P_2} a_i \Delta_i}{A_{P_1\cup P_2}}$. Also, $A_{P_1\cup
    P_2}=A_{P_1}+A_{P_2}$. Hence, $A_{P_1}<A_{P_1\cup P_2}$ and
    $A_{P_2}<A_{P_1\cup P_2}$. Therefore, 
    \begin{equation}
        \begin{aligned}
            \Delta(D_{P_1})+\Delta(D_{P_2})&=\frac{\sum_{p_i\in P_1} a_i\Delta_i}{A_{P_1}} + \frac{\sum_{p_i\in P_2} a_i\Delta_i}{A_{P_2}}\\
            &> \frac{\sum_{p_i\in P_1} a_i\Delta_i}{A_{P_1\cup P_2}} + \frac{\sum_{p_i\in P_2} a_i\Delta_i}{A_{P_1\cup P_2}}\\
            &=\frac{\sum_{p_i\in P_1\cup P_2} a_i\Delta_i}{A_{P_1\cup P_2}}\\
            &= \Delta(D_{P_1}+D_{P_2})
        \end{aligned}
    \end{equation}
\end{proof}

Now, we prove \Prop~\ref{prop:avg-inequality}.

\begin{proof}
    \begin{equation}
        \begin{aligned}
            \Delta_{D_P-D_{p_j}}&=\frac{\sum_{p_i\in P}a_i\Delta_i-a_j\Delta_j}{A_{D-p_j}}\\
            &=\frac{\sum_{p_i\in P}a_i\Delta_i-a_j\Delta_j}{A_{D}-a_j}\\
            &< \frac{\sum_{p_i\in P}a_i\Delta_i}{A_{D}}\\
            &=\Delta_{D_P}
        \end{aligned}
    \end{equation}
    The inequality in the above equation is true because
    $\frac{A}{B}<\frac{C}{D}$ implies $\frac{A-C}{B-D}<\frac{A}{B}$ where $A,B,C,D$
    are positive and $A>C$, $B>D$.
\end{proof}

\subsection{Generating Synthetic Data}

\parh{\ding{174} Synthetic Data A (SYN-A).}~We use Erdős-Rényi model, a
well-established random graph model, to synthesize random causal graphs of
different scales and to produce datasets via forward
sampling~\cite{poole2010artificial}. To simulate causally insufficient systems,
we mask $5\%$ variables at random and return the corresponding PAG (Partial
Ancestral Graph) as the ground truth. We construct conditional probability
tables based on a Dirichlet prior and generate two additional FD nodes on each
leaf node. Afterwards, these functional dependencies are employed to build
FD-induced graphs.

\parh{\ding{174} Synthetic Data B (SYN-B).}~In particular, we design a data
generating process with three variables $X,Y,Z$, where $X$ is a binary variable,
$Y$ is a categorical variable, and $Z$ is a numerical variable. Different values
in $X$ first impact $Y$ and then $Y$'s values further impact $Z$. When $Y$'s
realizations are equal to some specific values (i.e., $Y\in \{y_1,\cdots
y_k\}$), $Z$ would be sampled from a Gaussian distribution with a higher mean
$\mu^*$; otherwise, $Z$ is sampled from another Gaussian distribution with a
lower mean $\mu<\mu^*$. Note that the higher $k$, the harder \xplainer\ (and
other tools) to comprehensively identify all filters to form the correct
explanation. As a result, given different $X$, the aggregated result of $Z$
differs. Recall \D~\ref{def:diff} where $X$ and $Z$ form the context and the
target of a \diff. Then, we seek to extract the explanation from $Y$ and the way
$Y$ impacts $Z$ specified in the data generating process (i.e., $Y=y_1\lor
\cdots\lor Y=y_k$) constitutes the ground truth of this \diff. We concretize the
above data generating process with different parameters (e.g., cardinality and
conditional probability table of $Y$ and conditional distributions of $Z$) and
yield 18 datasets of different difficulties. By default, we generate
\textbf{SYN-B} datasets with 10,000 rows; the variable $Y$ contains ten values,
three of which would trigger abnormal $Z$ (i.e., $Z\sim \mathcal{N}(\mu^*,10),
\mu^*=60$) while the others would produce normal $Z\sim \mathcal{N}(\mu,10),
\mu=10$. The parameters are on a par with the configuration in Scorpion.

\end{document}